\documentclass[pdflatex,sn-mathphys-num]{sn-jnl}

\usepackage{graphicx}%
\usepackage{multirow}%
\usepackage{amsmath,amssymb,amsfonts}%
\usepackage{amsthm}%
\usepackage{mathrsfs}%
\usepackage[title]{appendix}%
\usepackage{xcolor}%
\usepackage{textcomp}%
\usepackage{manyfoot}%
\usepackage{booktabs}%
\usepackage{algorithm}%
\usepackage{algorithmicx}%
\usepackage{algpseudocode}%
\usepackage{listings}%

\begin{document}

\title{The Role of the Core in Setting Massive Neutron-Star Radii}

\author*[1]{\fnm{R.V.} \sur{Lobato}}
\email{lobato@cbpf.br}

\author[2]{\fnm{J.E.} \sur{Horvath}}
\email{foton@iag.usp.br}

\author[3]{\fnm{M.} \sur{Malheiro}}
\equalcont{In memoriam.}

\affil*[1]{\orgname{Centro Brasileiro de Pesquisas F\'isicas},
\orgaddress{\street{Rua Dr. Xavier Sigaud 150}, \city{Rio de Janeiro},
\postcode{22290-180}, \state{RJ}, \country{Brazil}}}

\affil[2]{\orgdiv{Instituto de Astronomia, Geof\'isica e Ci\^encias Atmosf\'ericas},
\orgname{Universidade de S\~ao Paulo},
\orgaddress{\city{S\~ao Paulo}, \state{SP}, \country{Brazil}}}

\affil[3]{\orgdiv{Departamento de F\'isica},
\orgname{Instituto Tecnol\'ogico de Aeron\'autica},
\orgaddress{\city{S\~ao Jos\'e dos Campos}, \state{SP}, \country{Brazil}}}

\abstract{
Recent observations by the Neutron Star Interior Composition Explorer
(NICER) indicate that the massive pulsar PSR~J0740+6620
($2.08\,M_{\odot}$) has a radius comparable to those inferred for
stars near $1.4\,M_{\odot}$, as exemplified by PSR~J0030+0451 and supported
by updated analyses. Such near-vertical mass-radius behavior is
difficult to obtain unless the high-density equation of state (EOS) stiffens
strongly. We show that, for massive pulsars, the leading radial scale can be
set by a relativistic, high-sound-velocity core beginning near twice the
nuclear saturation scale, while the outer low-density layer, comprising the
true crust and outer core, supplies only a subdominant correction. The key
evidence comes from radius decomposition: replacing the low-density branch
below the matching point with
unified EOS models spanning a factor of two in transition pressure changes the
$2.08\,M_{\odot}$ core radius by only $\sim160$~m, about half the variation of
the total radius. This convergence is genuinely a high-mass phenomenon: at
$1.4\,M_{\odot}$ the core radius remains more branch-sensitive, varying by
about $0.5$~km. Fixed-fraction comparisons confirm that core dominance
emerges systematically as the star grows more massive, where the traditional
separation between ``radius physics'' and
``maximum-mass physics'' breaks down. We establish this picture by combining
analytical Tolman~VII profiles and thin-crust matching, numerical
Tolman-Oppenheimer-Volkoff integration with a piecewise EOS, and direct-grid
Bayesian inference using NICER and GW170817 constraints under causal and
mass-support filters. Current data modestly constrain the
transition stiffness but leave the quadratic stiffening governed mainly by
causality. A comparison with a constant-sound-speed core remains
inconclusive: both descriptions produce comparable core-dominated radii,
leaving a degeneracy that sub-kilometre radius measurements can break.
}


\maketitle

\section{Introduction}
\label{sec:intro}

Understanding the relation between the equation of state (EOS) of dense matter and the observable structure of neutron stars remains one of the central problems in relativistic astrophysics. In recent years, simultaneous mass-radius measurements provided by the Neutron Star Interior Composition Explorer (NICER) have substantially refined this discussion by directly probing the stiffness of matter at supranuclear densities. In particular, the measurements of PSR J0030+0451 and the massive pulsar PSR J0740+6620 have introduced an intriguing structural constraint: despite their significantly different masses, both stars appear to possess comparable radii within current observational uncertainties \cite{riley/2019, miller/2019, riley/2021, miller/2021}. Updated reanalyses incorporating additional NICER exposure and XMM-Newton data have, if anything, sharpened this picture, inferring a radius for PSR J0740+6620 of $12.92^{+2.09}_{-1.13}$ km \cite{dittmann/2024} and $12.49^{+1.28}_{-0.88}$ km \cite{salmi/2024}, fully overlapping the updated radius inference $13.11\pm1.30$~km for the $1.34\,M_{\odot}$ pulsar PSR J0030+0451 \cite{vinciguerra/2024}.

This result challenges the intuitive expectation that a substantially heavier neutron star should undergo substantial radial compression. Standard discussions of neutron-star structure often emphasize that the stellar radius is primarily correlated with the pressure of matter at low and intermediate densities, while the maximum mass is determined by the extreme high-density core \cite{lattimer/2016}. Within that framework, the core prevents gravitational collapse, whereas the observable radius is comparatively less sensitive to the EOS at the highest densities. It has long been recognized that distinct density segments of the EOS govern different global observables: the radius of a $1.4\,M_{\odot}$ star is controlled largely by the pressure at one-to-three times saturation density, whereas the maximum mass reflects the EOS at substantially higher densities \cite{lattimer/2016, fortin/2015}. More recently, a systematic comparison of thin-crust approximations against exact TOV solutions has shown that this clean separation is only partially valid: the core radius itself introduces a radius uncertainty at least comparable to that arising from the treatment of the crust, so the stellar radius cannot be attributed to the subnuclear EOS alone \cite{kopp/2026}. Here we extend that observation into the massive-star regime, where the relative contributions of the crust and the core shift toward the latter as the main element setting the stellar radius.

The NICER observations therefore motivate a more nuanced, mass-dependent interpretation. For stars with masses near $1.4\,M_{\odot}$, the radius may indeed retain strong sensitivity to intermediate-density matter. Once the stellar mass approaches the $2\,M_{\odot}$ regime, however, the star probes much deeper into the EOS and the relativistic inner core changes the structural response. If the EOS stiffens sufficiently rapidly above approximately $2\epsilon_0$, where $\epsilon_0$ is the saturation energy-density scale, the high-density core can dominate the radius size while the crust remains a finite, subdominant contribution to the observable radius.

The same regime is probed from a different direction by the most massive pulsars known. The black-widow pulsar PSR J0952-0607, with $2.35\pm0.11\,M_{\odot}$~\cite{romani/2025}, lies well above the mass of PSR J0740+6620 and requires the high-density EOS to remain stiff enough to support configurations approaching $2.4\,M_{\odot}$, while inferences from the observed mass distribution point to still larger maximum masses~\cite{rocha/2023}. Any description of the massive-star radius must therefore be compatible with a maximum mass in this range.

In this picture, the near-equality of the radii inferred for PSR J0030+0451 and PSR J0740+6620 is not accidental. It reflects the onset of strong relativistic stiffening that reduces further radial compression as the stellar mass increases. The crust still contributes to the total radius, but for sufficiently massive stars it becomes a progressively smaller outer correction around a comparatively rigid high-density core. In the massive-star regime, the traditional separation between ``radius physics'' and ``maximum-mass physics'' therefore becomes increasingly blurred.

To investigate this possibility, we combine three complementary ingredients. First, we use the exact Tolman VII solution with the thin-crust approximation to obtain analytical insight into how the core radius enters the total stellar radius \cite{tolman/1939, zdunik/2008}. Second, we solve the Tolman-Oppenheimer-Volkoff (TOV) equations \cite{tolman/1939, oppenheimer/1939} numerically with piecewise equations of state in which a standard SLy4 low-density branch is matched to phenomenological relativistic core extensions above $2\epsilon_0$. Third, we infer the two parameters that control this high-density branch, namely the sound-speed squared at the transition, $s_{\rm tr}$, and the quadratic stiffening coefficient, $b$, with a Bayesian grid analysis combining NICER mass-radius measurements of PSR J0030+0451 and PSR J0740+6620 with GW170817 tidal-deformability information, while enforcing the causal condition $s_{\max}\le 1$ throughout each stellar sequence. This construction isolates the structural consequences of rapid high-density stiffening while maintaining causal propagation throughout the stellar interior.

We find that sufficiently stiff relativistic cores naturally generate near-vertical mass-radius sequences compatible with current NICER constraints. In these configurations, the high-density core spans a large fraction of the stellar radius and supplies the pressure support that controls the leading radius size of massive pulsars. The Bayesian posterior further shows that progressive stiffening is fully consistent with current multimessenger data, although the Bayes factor relative to a constant-sound-speed reference remains inconclusive to discriminate the cases. This suggests that multimessenger observations may already be approaching the precision needed to test a transition from crust/intermediate-density radius sensitivity in $1.4\,M_\odot$ stars to core-dominated radius size in the massive-star regime.

To confirm that this behavior is a property of the stiff core rather than of the particular low-density model, we further verify that the massive-star core radius is essentially unchanged when the sub-$2\epsilon_0$ branch is replaced by unified equations of state spanning a wide range of low-density stiffness, so that at $2.08\,M_{\odot}$ the core radius changes by only $\sim160$~m across branches spanning a factor of two in transition pressure, half the corresponding variation of the total radius.

\section{The Analytical Approach: Tolman VII and the Thin Crust}
\label{sec:analytical}

The relativistic thin-crust framework used here is well established. The matching relation between the core boundary and the total radius, in the form adopted below, was derived by Zdunik, Bejger, and Haensel \cite{zdunik/2008} and was recently validated against exact TOV integrations for a range of realistic equations of state by K\"opp, Horvath, and Vasconcellos \cite{kopp/2026}, who showed that it reproduces stellar radii to within a few hundred metres. Our purpose is therefore not to rederive the matching relation, but to expand it in the massive-star limit. Starting from the established equation, we obtain a closed expression for the fractional crustal correction that makes explicit why this correction shrinks as the stellar mass increases. This expansion is the new analytical content of the present work and provides the foundation for the numerical results that follow.

To analytically examine how the core sets the leading scale of the total stellar radius $R_\ast$ in massive pulsars, we adopt the Tolman VII solution \cite{tolman/1939} for the high-density core. Writing $r$ for the radial coordinate, we denote the boundary radius of this analytical thin-crust construction by $R_B$. The analytical core density profile $\rho_{\rm core}(r)$ is taken as
\begin{equation}
    \rho_{\rm core}(r) = \rho_c \left[ 1 - \left( 1 - \frac{\rho_t}{\rho_c} \right) \left( \frac{r}{R_B} \right)^2 \right],
    \label{eq:tolman7}
\end{equation}
where $\rho_c$ is the central mass density and $\rho_t$ is the mass density at the matching point. In the schematic limit one may identify $\rho_t$ with a density of order the nuclear saturation rest-mass density $\rho_0$, so that at $r=R_B$ the Tolman VII core transitions to the outer crust/envelope. The quadratic form is especially useful because it keeps the core profile analytic while preserving the physically expected monotonic decrease of the density from the center to the matching point.

To model the envelope atop this core, we adopt the standard relativistic thin-crust assumptions: pressure and enclosed mass are matched at $R_B$, the outer layer contributes negligibly to the mass, and the geometry and enclosed mass may be evaluated at the core boundary throughout the layer \cite{zdunik/2008, kopp/2026}. The low-density layer is represented by a BPS-like effective polytrope \cite{baym/1971}, $P=K\rho^{\gamma}$ with $\gamma$ close to $4/3$, following the usual crust estimates \cite{lattimer/2016}.
With these standard assumptions, the enthalpy matching relation of the relativistic thin-crust approximation is \cite{zdunik/2008, kopp/2026}
\begin{equation}
    \chi_t \equiv \int_0^{P_t} \frac{dP}{\epsilon + P}
    = \frac{1}{2}\ln\!\left[\frac{1-r_g/R_\ast}{1-r_g/R_B}\right],
    \label{eq:chi_t}
\end{equation}
where $R_\ast$ is the total stellar radius and $r_g \equiv 2GM/c^2$, with $G$ being Newton's gravitational constant and $c$ the speed of light. For the effective polytrope one has, to leading order in $P_t/\rho_t c^2$,
\begin{equation}
    \chi_t \simeq \frac{\gamma}{\gamma-1}\frac{P_t}{\rho_t c^2},
    \label{eq:chi_poly}
\end{equation}
where $\chi_t$ is the quantity denoted $\Theta$ in Ref.~\cite{kopp/2026}. For the BPS-like value $\gamma=4/3$, this becomes $\chi_t\simeq 4K\rho_t^{1/3}/c^2$.

We define $\alpha \equiv e^{2\chi_t}$; solving Eq.~\eqref{eq:chi_t} exactly for the total radius yields
\begin{equation}
    R_\ast = \frac{R_B}
    {1-(\alpha-1)\left(\frac{R_Bc^2}{2GM}-1\right)}.
    \label{eq:radius_core}
\end{equation}
With $\chi_t\equiv\Theta$ and $R_B$ identified with the boundary radius used in the thin-crust construction, Eq.~\eqref{eq:radius_core} coincides with the corresponding expression in Ref.~\cite{kopp/2026}. We now expand this relation in the massive-star, thin-crust limit. For $\chi_t\ll 1$, Eq.~\eqref{eq:radius_core} reduces to
\begin{equation}
    \Delta R_{\rm crust} \equiv R_\ast-R_B
    \simeq \chi_t\frac{R_B^2 c^2}{GM}
    \left(1-\frac{2GM}{R_Bc^2}\right).
    \label{eq:thincrust}
\end{equation}

Equation~\eqref{eq:thincrust} is central to the analytical argument because it shows why the crustal radial correction becomes progressively smaller in massive neutron stars. The first factor, $\chi_t$, is an enthalpy-like quantity fixed by the low-density crust EOS. Since $\chi_t\ll1$ for a thin crust, the envelope supplies only a modest radial correction once the core boundary and gravitational mass are specified. The second factor, $R_B^2/M$, suppresses the crustal extension as the stellar mass increases, especially in the massive-star regime where the stiff core radius varies only weakly with $M$. Finally, the relativistic factor $(1-2GM/R_Bc^2)$ further reduces the crustal contribution as the core compactness increases. Thus, as the star approaches the relativistic high-compactness regime, the outer layer is gravitationally compressed and contributes a diminishing, but finite, fraction of $R_\ast$.

The corresponding fractional correction is
\begin{equation}
    \frac{\Delta R_{\rm crust}}{R_B}
    \simeq \chi_t\frac{R_B c^2}{GM}
    \left(1-\frac{2GM}{R_Bc^2}\right).
    \label{eq:thincrust_fraction}
\end{equation}
Equations~\eqref{eq:thincrust} and \eqref{eq:thincrust_fraction} are therefore the desired massive-star expansion of Eq.~\eqref{eq:radius_core}. For a massive star with a stiff core, $R_B$ changes slowly while $M$ and the compactness increase; both effects drive $\Delta R_{\rm crust}/R_B$ downward. This result is independent of the detailed numerical implementation of the high-density EOS. It follows directly from relativistic structure arguments once the core is sufficiently compact and the crust remains a thin, low-enthalpy layer.

The Tolman VII profile also gives an explicit relation between the core boundary and the enclosed mass,
\begin{equation}
  M \simeq \frac{4\pi}{15}\left(2\rho_c+3\rho_t\right)R_B^3,
\end{equation}
so that
\begin{equation}
R_B = A M^{1/3}, \quad \text{with}\quad A = \left[\frac{15}{4\pi(2\rho_c+3\rho_t)}\right]^{1/3}.
    \label{eq:tolman_mass_scaling}
  \end{equation}
This is the Tolman VII core-radius scaling and is equivalent to the $R_{\rm core,thin}$ used in Ref.~\cite{kopp/2026} in the corresponding fixed-density limit. It also makes explicit that, at fixed characteristic core density, the factor $R_B^2/M$ in Eq.~\eqref{eq:thincrust} scales as $M^{-1/3}$.

It is worth making explicit where the choice of matching density enters
Eq.~\eqref{eq:thincrust}. The enthalpy integral $\chi_t$ depends on where the
matching is performed, but it is a constant once that point is fixed and
carries no dependence on the stellar mass. The entire mass dependence resides
in the geometric factor $R_B^2/M$ and in the relativistic factor
$(1-2GM/R_Bc^2)$, and the matching density enters the $M^{-1/3}$ scaling just
established only through the prefactor $A$. Changing the matching density
therefore rescales the amplitude of the outer correction without altering how
it scales with mass. The thinning of the outer layer along the massive-star
branch is thus a property of relativistic stellar structure at any fixed
density threshold, and not an artifact of the particular threshold adopted
here.

Substituting Eq.~\eqref{eq:tolman_mass_scaling} into Eq.~\eqref{eq:radius_core}, and using the $\gamma=4/3$ crustal estimate for $\chi_t$, gives the analytical radius scaling
\begin{equation}
    R_\ast = r_{g}\left[1 - \left(1 - \frac{2GM^{2/3}}{A c^2}\right)
    \exp\!\left(\frac{8K\rho_t^{1/3}}{c^2}\right)\right]^{-1}.
    \label{eq:radius_scaling}
\end{equation}
When $\rho_t$ is identified with a density of order $\rho_0$, Eq.~\eqref{eq:radius_scaling} reduces to the usual schematic thin-crust expression written in terms of the saturation density. This expression should be read as a fixed-characteristic-density toy limit: the coefficient $A$ is constant only if $\rho_c$ and $\rho_t$ are held fixed, or treated as representative core densities, while $M$ is varied. Along an actual one-parameter TOV sequence, the central energy density $\epsilon_c$ changes self-consistently with the stellar mass, so $A$ is not a universal constant for the full sequence. The analytical scaling nevertheless provides the relevant intuition: once a stiff core radius varies only weakly with mass, the crustal correction shrinks with compactness and near-vertical mass-radius behavior can emerge. This expectation is checked against the numerical sequence below.

\section{Numerical Methodology}
\label{sec:numerical}

Before specifying the numerical methodology, we fix the density convention used in the implementation. In the analytical Tolman-VII and thin-crust expressions, $\rho_c$, $\rho_t$, and $\rho$ denote mass densities; in the weakly relativistic crust estimate this is related to the energy density by $\epsilon\simeq\rho c^2$. In the numerical TOV calculation, $\epsilon$ denotes the total relativistic energy density entering the TOV equations. We adopt the saturation energy-density scale $\epsilon_0=150\,\mathrm{MeV\,fm^{-3}}$ and impose the matching point directly at $\epsilon_{\rm tr}=2\epsilon_0=300\,\mathrm{MeV\,fm^{-3}}$. The transition-density sensitivity analysis is therefore expressed throughout in terms of $\epsilon_{\rm tr}/\epsilon_0$.

To solve the Tolman-Oppenheimer-Volkoff (TOV) equations numerically, we construct a piecewise equation of state in which the low-density sector is supplied by the unified SLy4 model of Douchin and Haensel \cite{douchin/2001}. In practice, the tabulated SLy4 low-density branch is interpolated as $P(\epsilon)$ and used from the stellar surface up to the transition energy density
\begin{equation}
    \epsilon_{\rm tr}=2\epsilon_0=300\,\mathrm{MeV\,fm^{-3}}.
    \label{eq:rhotr}
\end{equation}
Although we loosely refer to this branch as the crust sector, numerically it provides the full low-density input required to connect the stellar surface to the onset of the stiff inner core. We therefore stress that the interface at $\epsilon_{\rm tr}=300\,\mathrm{MeV\,fm^{-3}}$ marks the onset of the stiff inner core in our construction and should not be confused with the standard crust-core transition, which occurs at $n_{cc}\simeq0.08$--$0.10\,\mathrm{fm}^{-3}$ \cite{kopp/2026}. The ``crust/envelope'' region shown in the figures and used in the radius decomposition of Sec.~\ref{sec:radius_decomp} therefore comprises the true crust \emph{plus} the SLy4 outer core. Correspondingly, we denote its thickness by $\Delta R_{\rm outer}$, reserving $\Delta R_{\rm crust}$ for the physical crust thickness obtained when the decomposition is performed at $n_{cc}$ \cite{kopp/2026}. The two decompositions are complementary: the former isolates the radial extent outside the imposed stiff high-density branch, whereas the latter isolates the solid crust.

We denote the radial coordinate of this imposed high-density interface by $R_{\rm core}$, so that the numerical outer-layer thickness is $\Delta R_{\rm outer}=R_\ast-R_{\rm core}$.

At the matching point we evaluate the SLy4 pressure at the fixed energy density,
\begin{equation}
    P_{\rm tr}=P_{\rm SLy4}(\epsilon_{\rm tr})
    =16.05\,\mathrm{MeV\,fm^{-3}},
    \label{eq:match_values}
\end{equation}
where $P_{\rm SLy4}(\epsilon)$ denotes the interpolated SLy4 pressure as a function of energy density. We require every high-density extension to satisfy
\begin{equation}
    \left. P_{\rm high}(\epsilon)\right|_{\epsilon=\epsilon_{\rm tr}} = P_{\rm tr},
    \qquad
    \epsilon \ge \epsilon_{\rm tr},
    \label{eq:matching_cond}
\end{equation}
where $P_{\rm high}$ denotes the pressure of the high-density continuation as a function of energy density. The piecewise EOS construction is intentionally phenomenological and is designed primarily to isolate the structural consequences of rapid relativistic stiffening above $\epsilon_{\rm tr}=2\epsilon_0$. At the matching point, pressure continuity is explicitly enforced, and the energy-density coordinate is continuous by construction because the high-density branch starts at $\epsilon_{\rm tr}$. This prescription maintains mechanical equilibrium and numerical stability during the TOV integration. The derivative $dP/d\epsilon$, equivalently the sound-speed squared $s=(c_s/c)^2$, is allowed to change discontinuously across the interface while $P$ and $\epsilon$ remain continuous. We therefore regard the matching as a phenomenological stiffness crossover in which the compressibility changes abruptly without introducing latent heat. It can represent the macroscopic pressure response associated with the onset of strongly interacting quark matter, quarkyonic matter, or other nonperturbative many-body repulsion \cite{mclerran/2019, annala/2020, annala/2023, fujimoto/2022}, but the discontinuity in $dP/d\epsilon$ alone does not determine the thermodynamic order of an underlying microscopic transition.

Consequently, the framework does not explicitly impose continuity of additional thermodynamic quantities such as the baryon chemical potential, baryon density, Gibbs free energy, or composition-dependent equilibrium conditions. A fully self-consistent microscopic treatment would require detailed modeling of nuclear composition, beta equilibrium, and interaction physics across the transition region. For the purpose of global stellar structure, however, the construction remains physically meaningful because the TOV equations depend directly on the relation $P(\epsilon)$, while the resulting EOS remains causal, mechanically stable, and observationally compatible with current neutron-star constraints. Future work should investigate thermodynamically complete matching procedures and smooth sound-speed interpolation schemes to determine how robust the core-dominated radius size remains under fully microscopic EOS treatments.

Above the matching density we consider two high-density extensions. The quadratic Tolman-VII-inspired EOS is our main model, while a linear constant-sound-speed EOS serves as a reference configuration against which the effect of the quadratic stiffening term can be measured.

\subsection{Linear Equation of State}
\label{sec:linear_eos}
A linear EOS assumes that the pressure increases proportionally with compression, with no additional stiffening mechanism and a sound speed that remains fixed throughout the core. We write this reference core EOS as
\begin{equation}
    P(\epsilon) = P_{\rm tr} + s\,(\epsilon-\epsilon_{\rm tr}),
    \qquad \epsilon \ge \epsilon_{\rm tr},
    \label{eq:linear_eos}
\end{equation}
where $c_s$ is the local sound speed and $s \equiv (c_s/c)^2$ is constant. Causality requires $0 < s \le 1$, and increasing $s$ makes the inner core stiffer, thereby enlarging both the maximum mass and the core radius. This parametrization is especially useful because it isolates the role of a relativistic sound speed above $2\epsilon_0$ with a single control parameter.

\subsection{Quadratic Tolman-VII-Inspired EOS}
\label{sec:quadratic_eos}
The quadratic Tolman-VII-inspired EOS generalizes the linear model by allowing the sound speed to grow dynamically with density. This produces progressive relativistic stiffening, stronger pressure support, reduced compressibility, and weaker radial contraction. Using the same matching point defined by Eq.~\eqref{eq:rhotr}, and therefore the same values $P_{\rm tr}$ and $\epsilon_{\rm tr}$, we write
\begin{equation}
    P(\epsilon)=P_{\rm tr}+s_{\rm tr}(\epsilon-\epsilon_{\rm tr})+b(\epsilon-\epsilon_{\rm tr})^2,
    \label{eq:quad_eos}
\end{equation}
with $P$, $\epsilon$, $P_{\rm tr}$, and $\epsilon_{\rm tr}$ expressed in $\mathrm{MeV\,fm^{-3}}$. Here $s_{\rm tr}\equiv (c_s/c)^2_{\rm tr}$ is the dimensionless sound-speed squared at the transition interface, while $b$ parametrizes progressively increasing repulsive interactions at supranuclear densities.

To integrate the TOV equations it is convenient to invert Eq.~\eqref{eq:quad_eos}, obtaining
\begin{equation}
    \epsilon(P) = \epsilon_{\rm tr} + \frac{-s_{\rm tr} + \sqrt{s_{\rm tr}^2 + 4b(P - P_{\rm tr})}}{2b}.
    \label{eq:quad_inv}
\end{equation}
For $b=0$ the analytic linear limit is recovered, $\epsilon(P)=\epsilon_{\rm tr}+(P-P_{\rm tr})/s_{\rm tr}$.
The corresponding dimensionless local sound-speed squared is
\begin{equation}
    s(\epsilon) \equiv \frac{dP}{d\epsilon} = s_{\rm tr} + 2b(\epsilon - \epsilon_{\rm tr}).
    \label{eq:quad_cs}
\end{equation}

In the modern dense-matter literature, a simple linear EOS with constant sound speed often cannot capture the complex dynamics of the inner core. The quadratic term governed by $b$ can effectively mimic several mechanisms known to trigger rapid high-density stiffening, including strong vector repulsion in relativistic mean-field or quark-matter descriptions, excluded-volume effects at extreme baryon density, higher-order many-body repulsive interactions, quarkyonic crossover stiffening, strongly coupled quark matter, and nonperturbative relativistic saturation effects \cite{kojo/2015, masuda/2013, mclerran/2019, typel/2016, tews/2018a}. In all of these scenarios, the pressure gradient with respect to energy density grows more rapidly than in a linear constant-sound-speed model. Macroscopically, this accelerated pressure response leads naturally to enhanced resistance against gravitational compression, larger maximum stellar masses, and a reduced radial contraction in the high-mass branch.

This behavior is closely connected to the energy-density-normalized trace anomaly $\Delta \equiv 1/3 - P/\epsilon$, which provides a dimensionless measure of conformality \cite{fujimoto/2022}. At low density $\Delta \simeq 1/3$, while conformal matter has $\Delta=0$; if $P/\epsilon$ temporarily exceeds $1/3$, $\Delta$ becomes negative. Its logarithmic derivative obeys $d\Delta/d\ln\epsilon=P/\epsilon-s(\epsilon)$, so a rapidly increasing sound speed drives a steep decline of $\Delta$ whenever $s(\epsilon)>P/\epsilon$. Fujimoto \textit{et al.} \cite{fujimoto/2022} showed that the sound-speed peak inferred for neutron-star matter arises from the derivative contribution of the trace anomaly, rather than simply from the sign of $\Delta$ itself. Bayesian analyses combining astrophysical data with perturbative-QCD input independently favor a sound speed that rises well above the conformal value at intermediate densities before softening in the cores of the most massive stars \cite{annala/2020, annala/2023}. Our linear-in-$\epsilon$ sound speed, Eq.~\eqref{eq:quad_cs}, is the simplest parametrization of the rising flank of this inferred profile across the density range probed by the configurations studied here: the single coefficient $b$ sets the rate at which $s(\epsilon)$ climbs above $s_{\rm tr}$ and therefore controls the corresponding decrease of $\Delta$. We do not attempt to model the subsequent post-peak softening, which lies beyond the central densities reached along our stable branch; our construction is therefore a controlled, causal representation of rapid high-density stiffening, not a complete sound-speed profile to asymptotic density.

We emphasize that the purpose of the present phenomenological construction is not to identify a unique microscopic origin for the stiffening. Because multiple distinct microphysical theories can yield similar macroscopic stiffness, our aim is instead to isolate the structural consequences of rapid high-density pressure growth in a model-independent manner. This interpretation connects directly to the analytical Tolman VII picture: the classical Tolman VII solution is characterized by a quadratic density profile, which suggests a slowly varying, spatially extended high-density core, while the quadratic EOS provides the dynamical pressure support required to realize and sustain such extended core configurations in full numerical solutions of the Tolman-Oppenheimer-Volkoff equations.
Within this framework, $b$ controls how rapidly the relativistic core transitions from moderately stiff to strongly stiff behavior as the central energy density increases. Equivalently, it controls the growth rate of the pressure response that counteracts the density gradient imposed by gravity. The limiting behaviors map directly onto the stellar structure:
\begin{itemize}
    \item \textbf{Small $b$ values:} The core behaves approximately like a linear EOS, generally producing softer compact stars that undergo standard radial contraction.
    \item \textbf{Intermediate to large $b$ values:} Progressive stiffening increasingly reduces radial contraction, supports the outer layers more efficiently, and raises the maximum mass $M_{\rm max}$.
    \item \textbf{Excessively large $b$ values:} Rapid pressure growth drives the local sound speed toward the causal limit, $s(\epsilon)=(c_s/c)^2\to 1$, marking the absolute physical boundary of allowable stiffening.
\end{itemize}
Causality therefore requires $s(\epsilon) \le 1$ throughout the star, placing an upper bound on $b$ once the highest central energy density to be explored is fixed. As an illustrative calibration aimed at reaching masses near $2.4\,M_{\odot}$ without violating causality, one may choose $s_{\rm tr}=0.40$ and $b=3.0\times10^{-4}\,(\mathrm{MeV\,fm^{-3}})^{-1}$ (now on called fiducial values); for the central energy density reached at the maximum mass, $\epsilon_c\simeq7.94\,\epsilon_0$, this yields $s_{\max}=0.935$.

The quadratic Tolman-VII-inspired EOS therefore provides a flexible and physically interpretable framework for studying how early relativistic stiffening above $2\epsilon_0$ can set the leading scale of the global stellar structure in massive stars. It demonstrates how a rapidly stiffening core can naturally produce the core-dominated radius size and near-vertical mass-radius branches suggested by recent NICER observations.

\subsection{Bayesian inference and numerical grid}
\label{sec:bayesian_method}

To determine whether the near-vertical mass-radius behavior requires a finely
tuned calibration, we perform a Bayesian inference in the $(s_{\rm tr},b)$
plane. The likelihood combines NICER mass-radius information for PSR
J0030+0451 and PSR J0740+6620 with the GW170817 tidal-deformability
constraint. Every EOS must satisfy three admissibility conditions:
$s_{\max}\le1$ at the central density of the maximum-mass configuration,
$M_{\max}\ge1.8\,M_\odot$, and a lowest mass on the sampled stable branch not
exceeding $1.34\,M_\odot$.

The two control parameters are the transition sound-speed squared,
\begin{equation}
    s_{\rm tr} \equiv \left(\frac{c_s}{c}\right)^2_{\rm tr},
\end{equation}
which controls the initial stiffness immediately above
$\epsilon_{\rm tr}=2\epsilon_0$, and the quadratic coefficient $b$, which
controls the progressive stiffening of the relativistic core at higher
densities. For each grid point $(s_{\rm tr},b)$, we construct the corresponding
piecewise EOS, integrate the TOV equations, compute the observables entering
the NICER and GW170817 likelihoods, and assign zero likelihood if any of the
three admissibility conditions is violated.

We assign priors that are uniform over the admissible region of each model.
For both models we take $s_{\rm tr} = (0.1-1)$; for the quadratic core we
additionally take
$b = (0-6\times10^{-4})\,{\rm MeV}^{-1}{\rm fm}^{3}$ before applying the
three filters above. The admissible region occupies $35.79\%$ of the
quadratic rectangle and $78.49\%$ of the linear interval. Each evidence is
normalized by the admissible volume of its own model. Fixing the prior volume
by the same physical conditions for both models, rather than normalizing one
model over an arbitrary enclosing interval, is essential for an unbiased
Bayes-factor comparison.

The likelihood incorporates the PSR J0740+6620 and PSR J0030+0451 radius
constraints through normalized two-piece normal distributions. For PSR
J0740+6620 we use $R_{2.08}=12.49$~km with
$\sigma_-=0.88$~km and $\sigma_+=1.28$~km from Salmi et al.
\cite{salmi/2024}; for PSR J0030+0451 we use
$R_{1.34}=13.11$~km with $\sigma_- = \sigma_+ =1.30$~km from Vinciguerra
et al. \cite{vinciguerra/2024}. Each two-piece density carries the
normalization $2/[\sqrt{2\pi}(\sigma_-+\sigma_+)]$. The PSR J0740+6620 mass
term is the cumulative probability of
$M_{\rm J0740}\sim\mathcal{N}(2.08,0.07^2)$ below the maximum mass supported
by each sequence, and its radius term is evaluated at $2.08\,M_\odot$. The
GW170817 tidal term is exactly flat for
$70\le\Lambda_{1.4}\le580$ and has Gaussian wings of width $50$ outside
that interval. The three likelihood factors are treated as independent, so
the mass-radius correlation present in the joint NICER posteriors is not
propagated; a fully consistent treatment would integrate over those
two-dimensional posteriors along each mass-radius sequence.

For each EOS, the central-energy-density sweep uses $100$ linearly spaced
points over $1.5\epsilon_0\le\epsilon_c\le10\epsilon_0$. The parameter plane
is integrated by direct quadrature on a uniform $180\times180$ grid, using
the trapezoidal rule with half weight on the grid boundaries. This yields the
Bayesian evidence and the joint posterior without stochastic sampling error;
the discretization tests reported below establish convergence to better than
$0.005$ in $\ln\mathcal{B}$.

\section{Results and Mass-Radius Profiles}
\label{sec:results}

\subsection{Mass-Radius Sequence and Internal Profiles}
\label{sec:mass_radius_profiles}

\begin{figure*}[h]
    \centering
     \includegraphics[width=\textwidth]{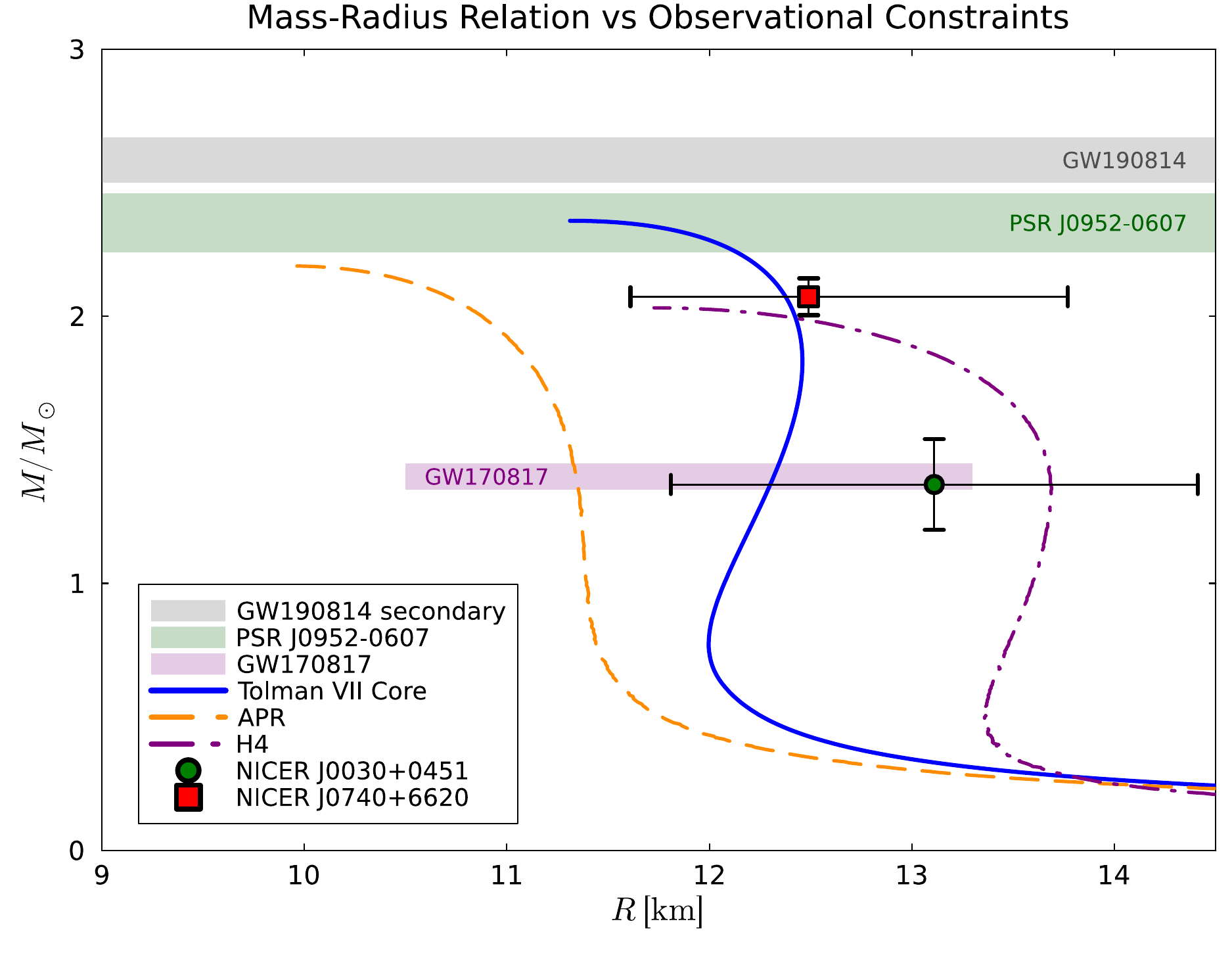}
    \caption{Mass-radius ($M$--$R_\ast$) comparison between standard literature EOS families and the present Tolman-VII-inspired stiff-core sequence. The Akmal-Pandharipande-Ravenhall (APR) and H4 curves were generated from EOS tables taken from the LALSuite development repository \cite{lalsuite}. The observational overlays show the GW170817 $1.4\,M_\odot$ radius band \cite{abbott/2018}, the green mass band for PSR J0952-0607 \cite{romani/2025}, the GW190814 secondary mass interval \cite{abbott/2020a}, and the NICER 68\% credible intervals for PSR J0030+0451 \cite{vinciguerra/2024} and PSR J0740+6620 \cite{salmi/2024}. APR exhibits stronger compactification, H4 remains larger-radius but left-bending, and the present model develops a near-vertical massive-star branch that overlaps the NICER constraints and reaches the PSR J0952-0607 mass band while remaining below the GW190814 mass interval.}
    \label{fig:MR}
\end{figure*}

Using the piecewise SLy4-quadratic EOS described above, we integrated the TOV equations from a central pressure $P_c$ outward to the surface condition $P = 0$. The resulting mass-radius ($M$--$R_\ast$) sequence, where $R_\ast$ denotes the total stellar radius in the numerical models, is shown in Fig.~\ref{fig:MR}.

The fiducial model generates a maximum mass of $M_{\max}=2.357\,M_{\odot}$, above the measured mass of PSR J0740+6620 ($2.08\pm0.07\,M_{\odot}$), consistent with PSR J0952-0607~\cite{romani/2025}, and below the GW190814 secondary mass interval. The fiducial curve should therefore be read as a massive-pulsar-compatible stiff-core benchmark rather than as an ultra-massive GW190814 solution. For a $1.4\,M_{\odot}$ pulsar, our model yields a radius of $R_\ast\approx12.32$~km, compatible with both the GW170817 band and the NICER radius interval for PSR J0030+0451. As the mass increases to $2.08\,M_{\odot}$ the radius remains essentially unchanged, $R_\ast\approx12.37$~km, varying by less than $0.1$~km over the entire range from $1.4$ to $2.08\,M_{\odot}$. This value lies close to the centre of the updated credible region inferred by Salmi \textit{et al.}~\cite{salmi/2024} and is likewise compatible with that of Dittmann \textit{et al.}~\cite{dittmann/2024}, so that the near-vertical trajectory is not obtained at the price of tension with either measurement.

This near-vertical trajectory in the $M$--$R_\ast$ plane supports the analytical premise: once the core transitions to a stiff relativistic EOS at $2\epsilon_0$, further structural compression can be substantially reduced. In this class of models for massive pulsars, $R_\ast$ is therefore not set by crustal physics alone; its leading scale and mass dependence are strongly affected by the pressure support and radial extent of the high-density core.

\begin{figure*}[h]
    \centering
    \includegraphics[width=\textwidth]{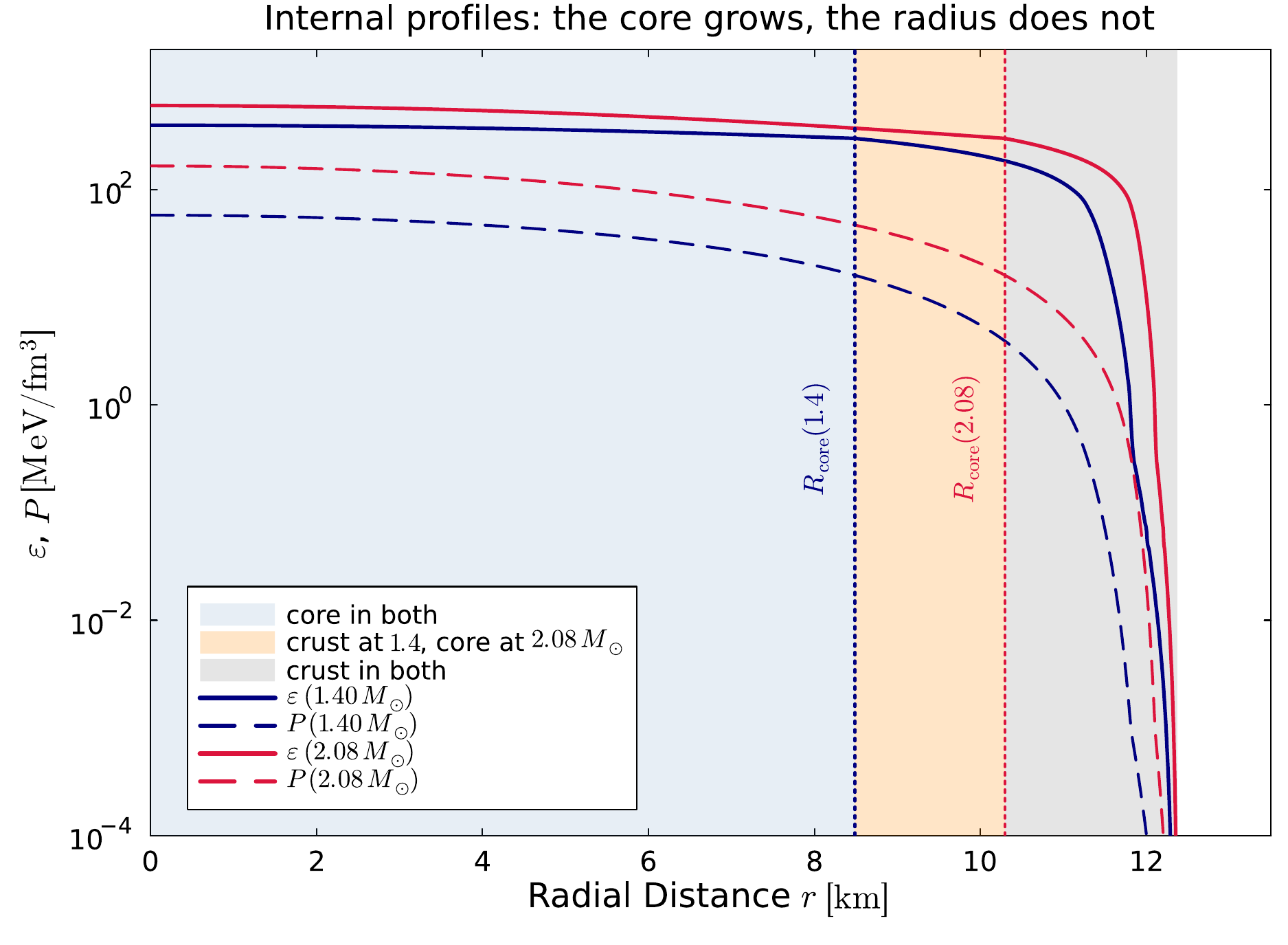}
    \caption{Internal energy-density and pressure profiles for the $1.40\,M_{\odot}$ and $2.08\,M_{\odot}$ stellar configurations. The blue and red dotted vertical lines indicate the corresponding values of $R_{\rm core}$, while the shaded regions distinguish the core shared by both configurations, the additional core region of the massive star, and their outer crust/envelope layers. The slow variation of $\epsilon$ and $P$ across the core, followed by their rapid decline in the outer layers, illustrates that the leading radial scale is controlled primarily by the stiff core, with the crust/envelope adding a finite outer extension.}
    \label{fig:profiles}
\end{figure*}

The same trend is visible in the internal profiles shown in Fig.~\ref{fig:profiles}. For the $2.08\,M_{\odot}$ configuration, both the energy density and pressure remain large and slowly varying throughout much of the stellar interior, indicating that the leading pressure-support scale is associated with a broad high-density core rather than with the low-density envelope. The vertical line marks the core radius, $R_{\rm core}$, while the shaded region denotes the crust/envelope. Across this outer layer the pressure falls rapidly toward zero and the energy density drops steeply only near the surface, consistent with a thin but finite radial correction to the total radius. The profile plot therefore provides a numerical counterpart to Eq.~\eqref{eq:thincrust}: once the core radius is fixed by the stiff high-density EOS, the observable stellar radius is shifted by a subdominant outer extension.

The role of the quadratic term is made explicit by the linear reference model of Sec.~\ref{sec:linear_eos}: with the same transition stiffness $s_{\rm tr}=0.40$ but $b=0$, the sequence reaches $M_{\max}=2.105\,M_{\odot}$, barely above the mass of PSR J0740+6620 and well below that of PSR J0952-0607, while the $1.4\,M_{\odot}$ radius is essentially unchanged, $R_{1.4}\approx12.30$~km. The progressive stiffening encoded by $b$ is therefore the ingredient that extends the stable branch to the highest observed pulsar masses while preserving the weak radius variation already allowed by a stiff constant-sound-speed core, leaving the low-mass sector essentially unchanged.

\subsection{Comparison with Standard Literature EOS Families}
\label{sec:eos_comparison}

The comparison with APR and H4 in Fig.~\ref{fig:MR} is included not merely to show that the present sequence can pass through the current observationally allowed regions. Its more important purpose is to determine whether the near-vertical mass-radius behavior obtained here represents a genuinely distinct structural regime relative to standard neutron-star EOS families. The APR and H4 comparison curves were generated using EOS tables taken from the LALSuite development repository \cite{lalsuite}, with the underlying APR and H4 models traced to the standard literature \cite{akmal/1998, lackey/2006, read/2009}. These two families provide useful reference cases because they span qualitatively different conventional responses: APR is comparatively compact, whereas H4 is substantially stiffer and produces larger radii.

The APR and H4 sequences should be regarded as representative benchmark implementations generated through the LALSuite EOS infrastructure. Minor implementation choices can shift the curves slightly, but this is secondary to the qualitative point: APR compactifies more strongly, H4 remains stiff but left-bending, and the Tolman-VII-inspired sequence shows substantially weaker radial contraction at high mass.

The qualitative behavior of the curves makes this distinction clear. The APR sequence exhibits the expected compactification trend, with the stellar radius decreasing noticeably as the mass increases. The H4 sequence, although much stiffer and shifted toward larger radii, still follows a conventional left-bending mass-radius trajectory as the star approaches the high-mass branch. By contrast, the Tolman-VII-inspired stiff-core model developed here produces a much more vertical branch in the massive-star regime: the radius changes only weakly while the gravitational mass increases toward the $2\,M_{\odot}$ scale.

This difference has a direct physical interpretation. In conventional EOS families, increasing mass is accompanied by stronger gravitational compression and progressive radius reduction. In the present model, however, the rapid relativistic stiffening above approximately $2\epsilon_0$ reduces strong radial contraction. The high-density core therefore behaves as a comparatively rigid pressure-supporting region, maintaining an extended core radius even as the star moves deeper into the relativistic regime.

The comparison also shows that the near-vertical branch is not simply a generic consequence of ``stiffness'' alone. Even a relatively stiff conventional EOS such as H4 continues to compactify appreciably once the mass increases. The crucial ingredient is therefore not merely large pressure support, but sufficiently early and rapid high-density stiffening capable of maintaining an extended relativistic core over the massive-star portion of the sequence.

This result connects directly to the central thesis of the paper. Standard EOS families such as APR and H4 remain broadly consistent with the traditional picture in which radius sensitivity is associated mainly with low- and intermediate-density matter. The present stiff-core regime enters a qualitatively different structural behavior: the relativistic core increasingly controls the radius size of massive stars. The precise locations of the comparison curves can depend on the chosen EOS implementation, crust matching, and interpolation scheme, but the global distinction is robust. The Tolman-VII-inspired sequence exhibits substantially weaker compactification than conventional APR- and H4-like EOS families, strengthening the interpretation that NICER observations may already be probing the onset of a core-dominated structural regime in sufficiently massive neutron stars.

\subsection{Dynamical Stability}
\label{sec:dynamical_stability}

\begin{figure*}[h]
    \centering
    \includegraphics[width=\textwidth]{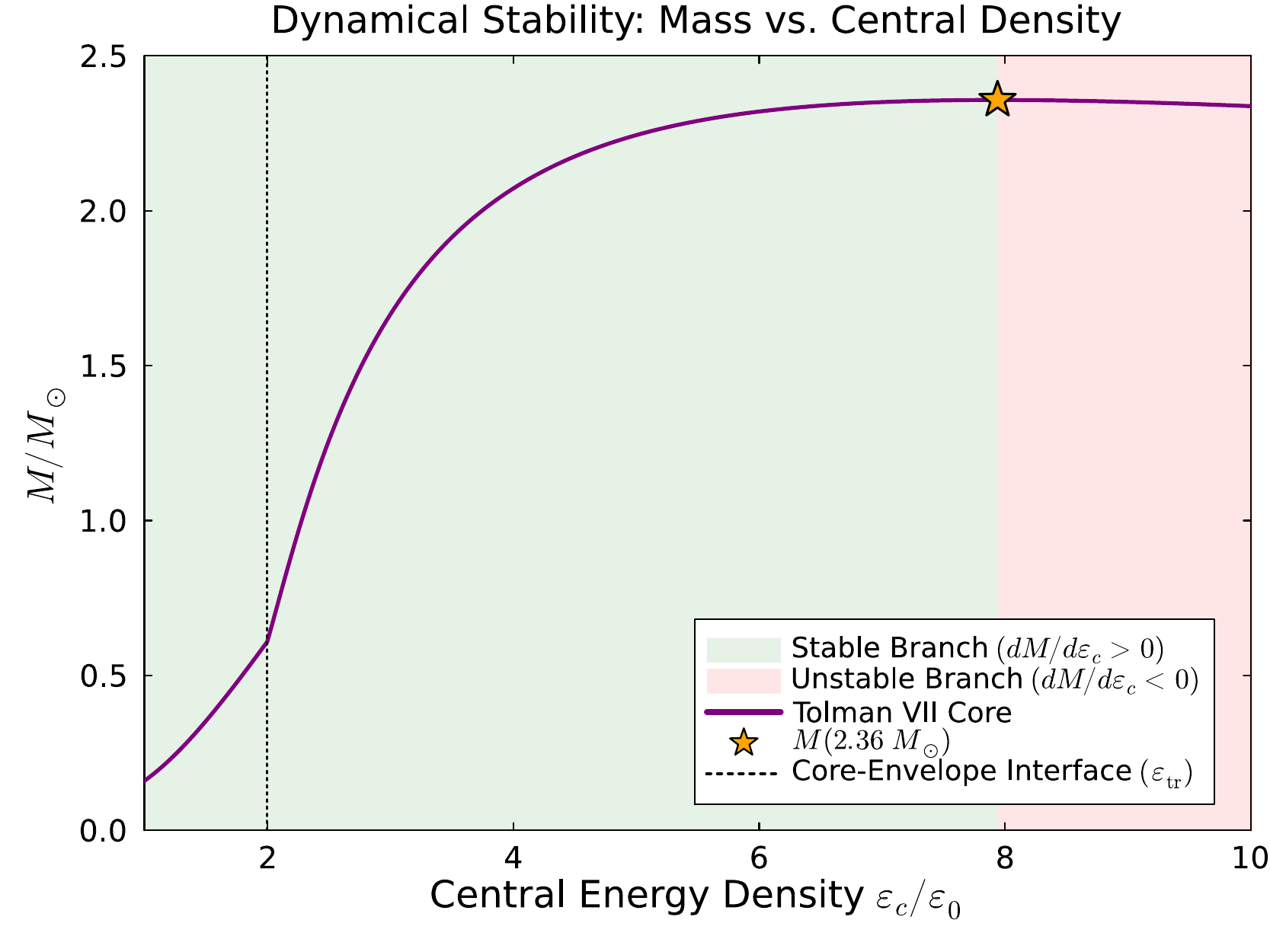}
    \caption{Gravitational mass as a function of central energy density for the Tolman-VII-motivated sequence. The dotted vertical line marks the core-envelope matching energy density, $\epsilon_{\rm tr}=2\epsilon_0$, on the central-density axis; configurations to its right contain the imposed high-density core. The green region satisfies $dM/d\epsilon_c>0$ and represents the stable branch by the standard turning-point criterion, while the red region beyond the maximum mass has $dM/d\epsilon_c<0$ and is unstable. The orange star marks $M_{\max}=2.357\,M_{\odot}$ at $\epsilon_c\simeq7.94\,\epsilon_0$, showing that the $2.08\,M_{\odot}$ configuration remains on the stable branch.}
    \label{fig:mrho}
\end{figure*}

Figure~\ref{fig:mrho} assesses dynamical stability by plotting the gravitational mass as a function of central energy density, normalized to the saturation scale. The dotted vertical line marks the core-envelope matching energy density used in the EOS construction, $2\epsilon_0$, shown on the central-density axis. Configurations to the right of this marker have central densities above the matching scale and therefore contain the Tolman-VII-inspired high-density core. In the Chandrasekhar radial-oscillation formulation, stability requires the squared frequency of the fundamental radial mode to remain positive, $\omega_0^2>0$; the onset of instability occurs when the mode becomes neutral, $\omega_0^2=0$ \cite{chandrasekhar/1964, friedman/1988}. For a one-parameter, nonrotating sequence at fixed EOS, the turning-point theorem identifies this neutral point with the mass extremum, $dM/d\epsilon_c=0$.

Along the green-shaded branch $dM/d\epsilon_c>0$, so the configurations are expected to have $\omega_0^2>0$. After the central density crosses the $\epsilon_{\rm tr}=2\epsilon_0$ matching marker, the sequence develops the Tolman-VII-inspired stiff core and remains stable up to the orange star, where $M_{\max}=2.357\,M_{\odot}$ near $\epsilon_c\simeq7.94\,\epsilon_0$. Beyond this point the red-shaded branch has $dM/d\epsilon_c<0$, indicating that the fundamental radial mode has acquired $\omega_0^2<0$ and that further compression cannot generate a stable equilibrium. The $2.08\,M_{\odot}$ model therefore lies on the stable side of the maximum-mass point.

\subsection{Compactness Evolution}
\label{sec:compactness}

\begin{figure*}[h]
    \centering
    \includegraphics[width=\textwidth]{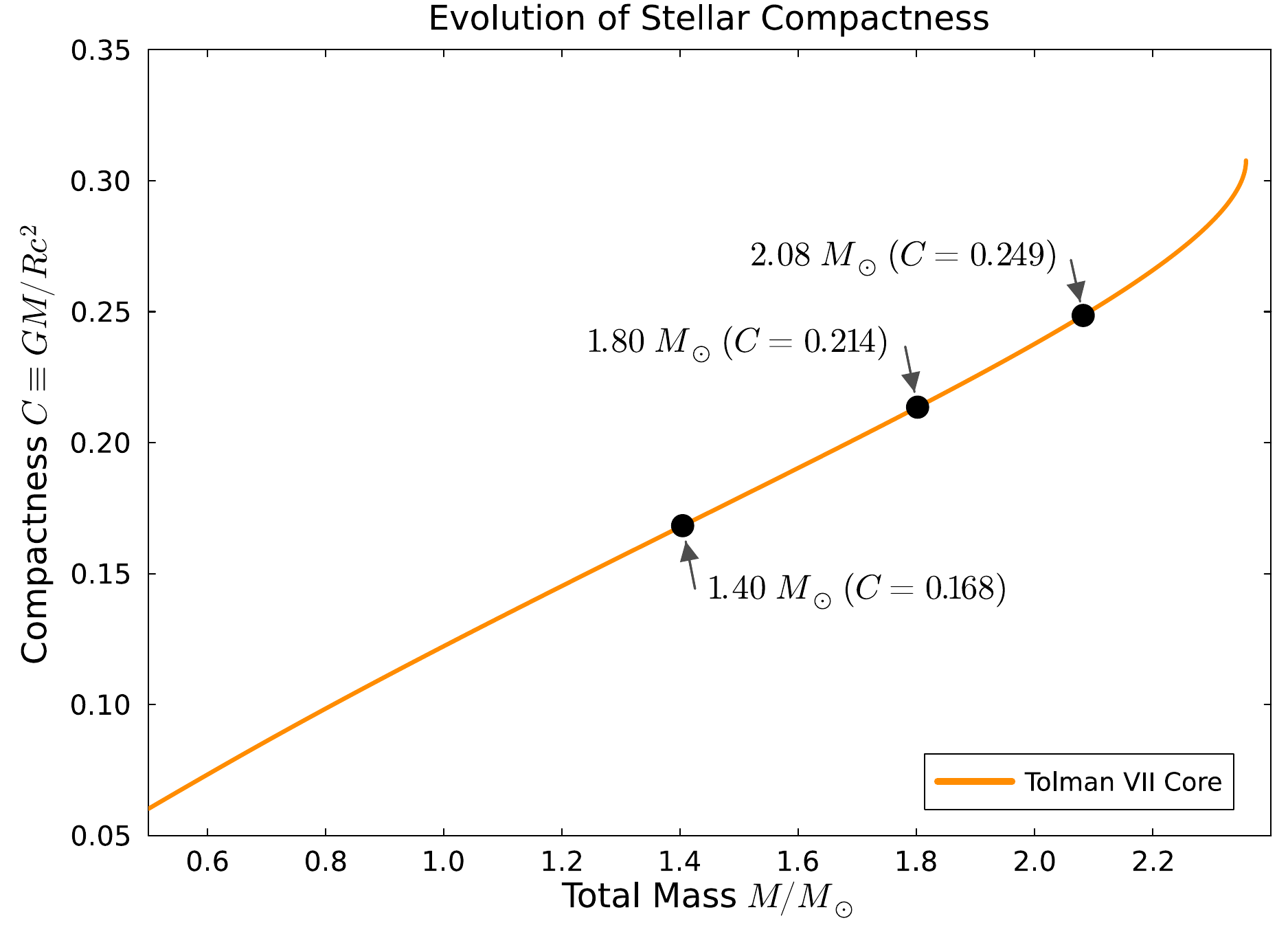}
    \caption{Evolution of the stellar compactness $C \equiv GM/(R_\ast c^2)$ as a function of total stellar mass $M$. The annotated points correspond to $1.40\,M_\odot$, $1.80\,M_\odot$, and the massive PSR J0740+6620 at $2.08\,M_\odot$, with $C=0.168$, $0.214$, and $0.249$, respectively. The smooth monotonic rise reflects the nearly constant total radius $R_\ast$ associated with strong relativistic core pressure support.}
    \label{fig:compactness}
\end{figure*}

The evolution of the stellar compactness along the stable branch provides another way to interpret the near-vertical mass-radius behavior. We define the dimensionless compactness as
\begin{equation}
    C \equiv \frac{GM}{R_\ast c^2}.
\end{equation}
Using the numerical TOV solutions generated for our piecewise SLy4-Quadratic EOS, we computed $C(M)$ from low-mass configurations up to the maximum-mass limit. The resulting compactness evolution is shown in Fig.~\ref{fig:compactness}.

For the three representative configurations highlighted in the figure, the compactness values are
\begin{equation}
\begin{aligned}
    C(1.40\,M_\odot) &= 0.168, \\
    C(1.80\,M_\odot) &= 0.214, \\
    C(2.08\,M_\odot) &= 0.249.
\end{aligned}
\label{eq:compactness_values}
\end{equation}
These values increase smoothly and monotonically along the stable branch. However, they do not show the rapid compactification typically associated with a strongly compressible or soft high-density EOS, where increasing mass is accompanied by a significant decrease in radius and hence by an accelerated rise in $C$.

In contrast, our stiff-core construction alters this response. The rapid quadratic increase in pressure above $\sim2\epsilon_0$ counteracts gravitational compression, so for masses $M \gtrsim 1.4\,M_{\odot}$ the rise in $C(M)$ is driven almost entirely by the increasing mass. Between $1.40$ and $2.08\,M_{\odot}$ the radius varies by less than $0.15$~km, passing through a shallow maximum near $1.8\,M_{\odot}$ ($R_\ast = 12.32$, $12.46$ and $12.37$~km at $1.40$, $1.80$ and $2.08\,M_{\odot}$ respectively), so that the $48\%$ increase in compactness over this interval reflects the $49\%$ increase in mass almost one to one. The $2.08\,M_{\odot}$ configuration already probes the relativistic high-density regime, but its compactness $C=0.249$ remains below the extreme causality-sensitive range. This behavior reinforces the interpretation that the massive-star sequence is supported by a stiff core that sets the leading radial scale, with the crust contributing a compressed outer correction.

This compactness evolution provides an additional structural diagnostic of the same mechanism. For sufficiently massive neutron stars, the observable radius becomes increasingly controlled by the relativistic high-density core, with the crust contributing a comparatively small and diminishing radial correction.

\subsection{Core Contribution and Radius Decomposition}
\label{sec:radius_decomp}

We now make the radial partition explicit. For massive neutron stars in this model, the radius size is increasingly governed by the high-density relativistic core, while the outer low-density layer contributes a smaller but finite correction. We therefore decompose the total radius and the associated geometric volume fraction. These quantities are structural diagnostics of the radial partition, not stand-alone measures of the overall stellar response.

We define the outer-layer radial contribution explicitly as:
\begin{equation}
\Delta R_{\rm outer} \equiv R_\ast - R_{\rm core},
\label{eq:radius_decomposition}
\end{equation}
where $R_\ast$ is the total observable stellar radius and $R_{\rm core}$ is the radial coordinate at the core-envelope interface, defined in our model by the fixed threshold $\epsilon_{\rm tr}=2\epsilon_0=300\,\mathrm{MeV\,fm^{-3}}$.

Using the numerical TOV solutions computed with the piecewise SLy4-Quadratic EOS, we evaluated $R_{\rm core}$, $R_\ast$, $\Delta R_{\rm outer}$, and the corresponding radial and geometric-volume core fractions at four representative masses. The results, summarized in Table \ref{tab:radii}, should be interpreted together with the pressure, energy-density, and compactness diagnostics discussed above.

\begin{table*}[t]
\caption{Core and outer-layer structural decomposition for the piecewise SLy4-Quadratic model ($s_{\rm tr} = 0.40$, $b = 3.0\times10^{-4}\,(\mathrm{MeV\,fm^{-3}})^{-1}$). The transition is fixed at $\epsilon_{\rm tr}=2\epsilon_0$ and $P_{\rm tr}=P_{\rm SLy4}(\epsilon_{\rm tr})$. As the total mass increases, the outer-layer radial thickness decreases and the geometric core fraction increases; these geometric diagnostics complement the pressure and compactness evidence that the stiff core sets the leading radius scale in massive configurations.}
\centering
\setlength{\tabcolsep}{6pt}
\renewcommand{\arraystretch}{1.2}
\begin{tabular}{c c c c c c}
\hline\hline
$M$ [$M_\odot$] & $R_\ast$ [km] & $R_{\rm core}$ [km] & $\Delta R_{\rm outer}$ [km] & $R_{\rm core}/R_\ast$ & $(R_{\rm core}/R_\ast)^3$ \\
\hline
1.00 & 12.07 & 6.45  & 5.62 & 0.53 & 0.15 \\
1.40 & 12.32 & 8.47  & 3.85 & 0.69 & 0.33 \\
2.08 & 12.37 & 10.29 & 2.08 & 0.83 & 0.58 \\
2.18 & 12.25 & 10.41 & 1.84 & 0.85 & 0.61 \\
\hline\hline
\end{tabular}
\label{tab:radii}
\end{table*}

The TOV integrations show that, for sufficiently massive neutron stars in this model, the macroscopic radius follows the response of the dense relativistic core rather than the low-density envelope alone. This inference is not based on radial or volume fractions by themselves; it also relies on the weak radius variation along the massive-star branch, the large relativistic pressure support in the core, and the compactness evolution discussed in Sec.~\ref{sec:compactness}. As detailed in Table \ref{tab:radii}, the geometric core fraction increases substantially with mass, providing an illustrative decomposition of the radial structure.

For a $1.4\,M_\odot$ star, the core spans approximately $69\%$ of the stellar radius, corresponding to a geometric volume fraction, $(R_{\rm core}/R_\ast)^3$, of roughly $33\%$. As the star is compressed to support a mass of $2.08\,M_\odot$, the outer-layer radial thickness decreases to $\Delta R_{\rm outer}\approx2.08$~km. The radial core fraction then rises to $83\%$, and the geometric core fraction rises to $58\%$. These fractions are illustrative structural diagnostics and should not be read as definitive evidence that geometry alone determines the stellar response. Together with the pressure and energy-density profiles in Fig.~\ref{fig:profiles}, the compactness evolution in Fig.~\ref{fig:compactness}, and the weak radius variation along the high-mass branch, they indicate that the stiff relativistic core increasingly governs the response to gravitational compression and sets the leading scale of $R_\ast$ in massive configurations. The true crust remains structurally present within the outer low-density layer, whose radial contribution becomes subdominant as compactness increases.

\subsection{Numerical validation of the thin-crust relation}
\label{sec:thin_crust_validation}

Having introduced the TOV sequence and the $2\epsilon_0$ core/envelope decomposition above, we now validate the analytical matching relation in the regime for which it is designed. Because Eqs.~\eqref{eq:radius_core}--\eqref{eq:thincrust_fraction} rest on the thin-crust assumption $\chi_t\ll1$, the check must be performed at the physical crust-core boundary rather than at the $2\epsilon_0$ interface used for the core-fraction diagnostic of Sec.~\ref{sec:radius_decomp}. We therefore evaluate the matching relation at $n_{\rm cc}\simeq0.076\,\mathrm{fm^{-3}}$ ($\epsilon_{\rm cc}\simeq0.48\,\epsilon_0$), the SLy4 crust-core transition, where the enthalpy integral of Eq.~\eqref{eq:chi_t} is $\chi_t=\int_0^{P_{\rm cc}}\!dP/(\epsilon+P)=0.024$, comfortably in the thin-layer regime. Taking $R_B$ as the TOV radial coordinate at $n_{\rm cc}$ and $\Delta R_{\rm thin}\equiv R_\ast-R_B$ as the numerically integrated crust thickness, Table~\ref{tab:eq5} compares the exact relation Eq.~\eqref{eq:radius_core} and its massive-star expansion Eq.~\eqref{eq:thincrust} against the TOV result along the fiducial SLy4-quadratic sequence described above.

The amplitude does matter for the accuracy of the expansion itself, however.
At the physical crust-core transition the enthalpy integral is
$\chi_t=0.024$, whereas at the $2\epsilon_0$ interface it is $0.097$, four
times larger, so the first-order truncation of $\alpha=e^{2\chi_t}$ that leads
to Eq.~\eqref{eq:thincrust} carries a relative error of order ten per cent
there against a few per cent at $n_{\rm cc}$. This is why the quantitative
validation is performed at the physical boundary, while the $2\epsilon_0$
decomposition of Sec.~\ref{sec:radius_decomp} is used only as a structural
diagnostic and never as an input to the analytical relation.

\begin{table*}[t]
\centering
\caption{Validation of the analytical matching relation at the crust-core boundary $n_{\rm cc}=0.076\,\mathrm{fm^{-3}}$ ($\chi_t=0.024$). Columns: gravitational mass; TOV total radius; core-boundary radius $R_B$; compactness parameter $2GM/R_Bc^2$; numerically integrated crust thickness $\Delta R_{\rm thin}$; the exact prediction of Eq.~\eqref{eq:radius_core}; and the leading-order expansion of Eq.~\eqref{eq:thincrust}. The exact relation reproduces the TOV thickness to better than $1\%$; the expansion is accurate to $\lesssim15\%$ and improves monotonically with mass as the crust thins, i.e.\ precisely in the massive-star regime of interest.}
\label{tab:eq5}
\begin{tabular}{ccccccc}
\toprule
$M$ & $R_\ast$ & $R_B$ & $\dfrac{2GM}{R_Bc^2}$ &
$\Delta R_{\rm thin}$ & Eq.~\eqref{eq:radius_core} & Eq.~\eqref{eq:thincrust} \\
$[M_\odot]$ & [km] & [km] & & [km] & [km] & [km] \\
\midrule
1.00 & 12.08 & 10.55 & 0.281 & 1.526 & 1.516 & 1.294 \\
1.40 & 12.32 & 11.27 & 0.368 & 1.046 & 1.040 & 0.930 \\
1.80 & 12.46 & 11.72 & 0.454 & 0.738 & 0.736 & 0.676 \\
2.08 & 12.37 & 11.81 & 0.521 & 0.561 & 0.559 & 0.521 \\
\bottomrule
\end{tabular}
\end{table*}

Two conclusions follow. First, the exact matching relation Eq.~\eqref{eq:radius_core} reproduces the numerically integrated crust thickness to better than $1\%$ across the full mass range, the agreement improving from $0.7\%$ at $1.0\,M_\odot$ to $0.4\%$ at $2.08\,M_\odot$, confirming that the relativistic thin-crust framework, validated against TOV integrations for realistic EOS by K\"opp \textit{et al.}~\cite{kopp/2026}, applies to the present piecewise construction. Second, the leading-order expansion Eq.~\eqref{eq:thincrust}, on which the qualitative massive-star argument of Sec.~\ref{sec:analytical} rests, agrees with the exact result to $\lesssim15\%$ and, importantly, converges toward it as the mass increases ($15\%$ at $1.0\,M_\odot$ to $7\%$ at $2.08\,M_\odot$): the higher-order terms in $\chi_t$, whose coefficients depend on the compactness, become progressively smaller as the crust thins on the massive-star branch. The three-factor suppression in Eq.~\eqref{eq:thincrust}, the enthalpy $\chi_t$, the geometric factor $R_B^2/M$, and the relativistic factor $(1-2GM/R_Bc^2)$; therefore captures both the magnitude and the mass-scaling of the crustal correction in the regime where it is applied.

We stress that this thin-crust decomposition, taken at $n_{\rm cc}$, is distinct from and complementary to the core/envelope decomposition of Sec.~\ref{sec:radius_decomp}, which is taken at the imposed high-density interface $\epsilon_{\rm tr}=2\epsilon_0$. The former isolates the true solid crust, for which the analytical relation is designed and against which it is validated here; the latter isolates the radial extent lying outside the stiff high-density branch, and its larger $\Delta R_{\rm outer}$ (Table~\ref{tab:radii}) accordingly includes the SLy4 outer core; at $2.08\,M_\odot$ the two decompositions give $0.561$~km and $2.08$~km, respectively. The analytical relation is not expected to reproduce the $2\epsilon_0$ thickness, and Table~\ref{tab:radii} is obtained directly from the TOV integration, not from Eq.~\eqref{eq:thincrust}.

The two boundaries are separated by the SLy4 outer core, which extends from
$n_{\rm cc}$ to $\epsilon_{\rm tr}$ and is by no means negligible: at
$2.08\,M_\odot$ it accounts for about $1.5$~km of the $2.08$~km outer layer,
roughly three times the physical crust thickness. The claim that the outer
region is subdominant therefore refers to the layer as a whole relative to the
stiff core, and not to the crust alone.

\subsection{Insensitivity to the Low-Density Branch}
\label{subsec:insensitivity}

The radius decomposition of Sec.~\ref{sec:radius_decomp} was carried out with a single low-density model (SLy4). A natural objection is that the near-vertical massive-star branch, and the core-dominance interpretation drawn from it, might be an artifact of that particular sub-$2\epsilon_0$ input rather than of the imposed stiff core. To test this, we replace the low-density branch below the fixed threshold $\epsilon_{\rm tr}=2\epsilon_0$ while holding the high-density stiffness parameters fixed at $(s_{\rm tr},b)=(0.40,\,3.0\times10^{-4}\,{\rm MeV}^{-1}{\rm fm}^{3})$. Pressure continuity is enforced separately for each branch, so $P_{\rm tr}$ is branch dependent while $\epsilon_{\rm tr}$ remains fixed: the continuations share the same derivatives controlled by $(s_{\rm tr},b)$ but are not identical pressure offsets. We consider five alternative unified branches drawn from the LALSuite tabulated set~\cite{lalsuite}: the Brussels-Montreal family BSk19, BSk20, and BSk21~\cite{goriely/2010,potekhin/2013}, constructed to span soft, intermediate, and stiff neutron-matter behavior, and two Skyrme models, KDE0v~\cite{agrawal/2005} and SKI4~\cite{reinhard/1995}. Across this set the matching pressure at the interface ranges from $7.3$ to $16.1$~MeV~fm$^{-3}$, a factor of $2.2$; the SLy4 baseline is the stiffest of the group, so our fiducial radii correspond to the large-$P_{\rm tr}$ end of the plausible range.

\begin{figure*}[h]
    \centering
    \includegraphics[width=\textwidth]{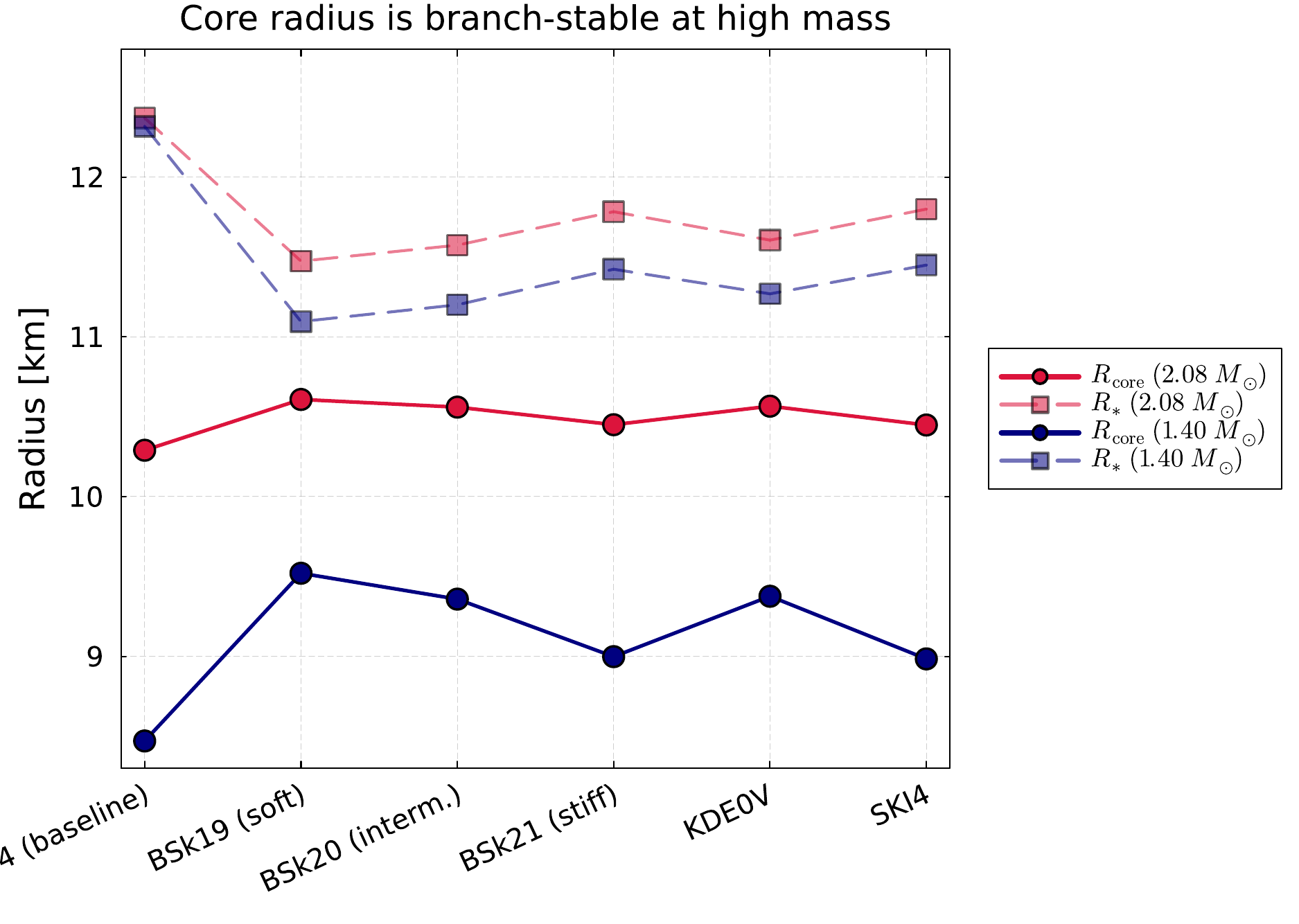}
    \caption{Fixed-mass core and total radii for a quadratic core with $s_{\rm tr}=0.40$ and $b=3.0\times10^{-4}\,\mathrm{MeV}^{-1}\mathrm{fm}^{3}$ matched at $\epsilon_{\rm tr}=2\epsilon_0$ to six low-density branches. Red curves correspond to $2.08\,M_\odot$ and blue curves to $1.40\,M_\odot$; circles and solid lines show $R_{\rm core}$, while squares and dashed lines show $R_\ast$. Across the five non-SLy4 branches, the $R_{\rm core}$ spread decreases from $536$~m at $1.40\,M_\odot$ to $160$~m at $2.08\,M_\odot$, whereas the corresponding total-radius spreads are $354$ and $326$~m. Including the stiff SLy4 baseline increases the $2.08\,M_\odot$ core-radius spread to $318$~m and shifts the total radius upward through its larger $\Delta R_{\rm outer}$.}
    \label{fig:core_universality}
\end{figure*}

The result is contained in the decomposition of the total radius into the core radius $R_{\rm core}$ and the outer contribution $\Delta R_{\rm outer}=R_\ast-R_{\rm core}$ of Eq.~\eqref{eq:radius_decomposition}, now evaluated branch by branch at fixed mass and summarized in Fig.~\ref{fig:core_universality}. At $M=1.40\,M_\odot$ the core radius varies by $536$~m across the five alternative branches, more than the $354$~m spread of the total radius: there the core and the outer layer are anticorrelated and partially compensate, a stiffer low-density branch yielding a smaller core radius but a correspondingly thicker envelope. At $M=2.08\,M_\odot$ this relation inverts. The core-radius spread falls to $160$~m, half the $326$~m spread of the total radius, so that the residual variation is carried mainly by $\Delta R_{\rm outer}$, which ranges from $0.87$ to $1.35$~km. The core has changed from being the branch-sensitive element to being the stable one, and the low-density branch is left setting the thickness of the outer layer rather than the leading radial scale.

\begin{figure*}[h]
    \centering
    \includegraphics[width=\textwidth]{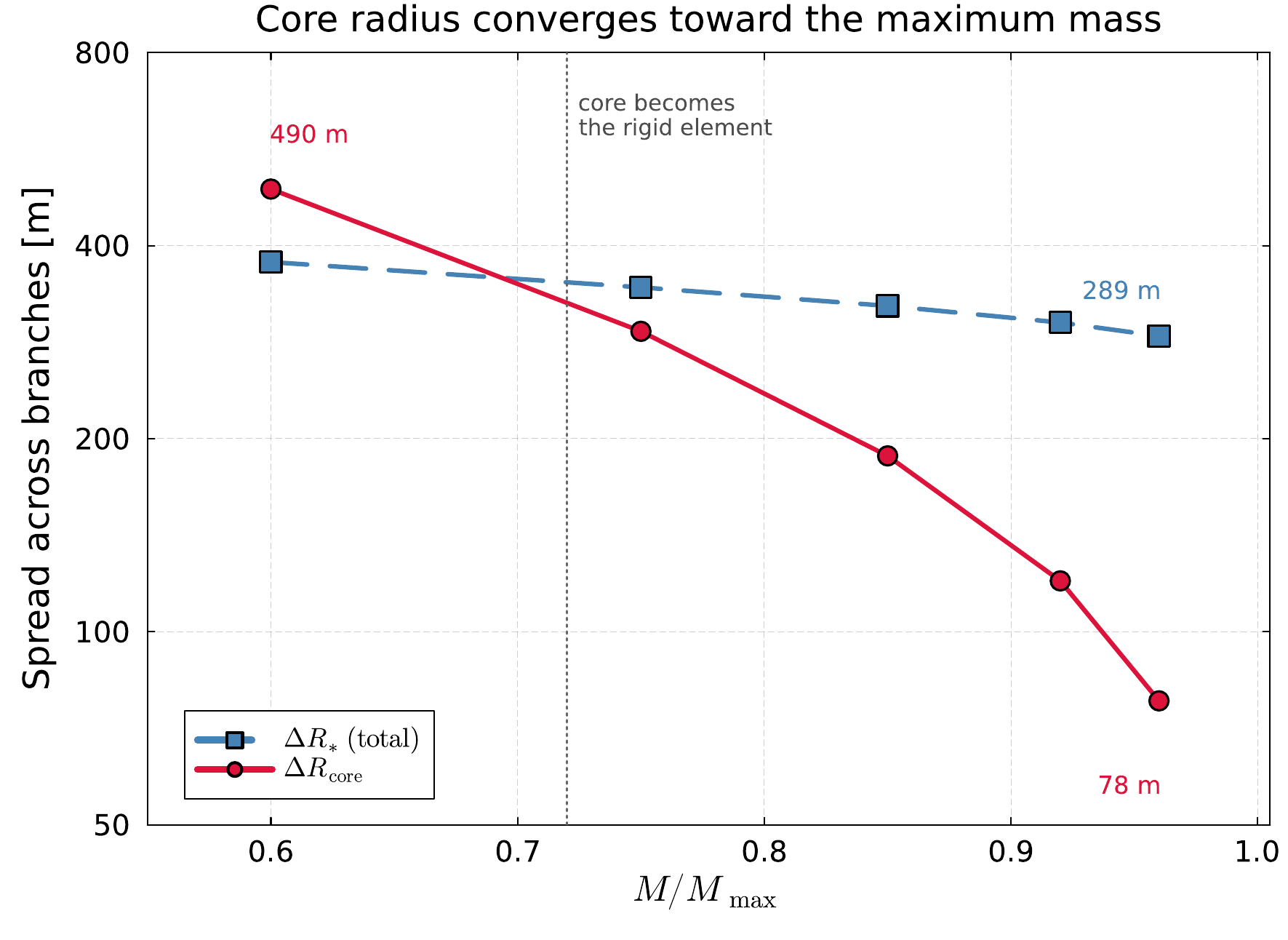}
    \caption{Spread of the core radius (red solid curve) and total radius (blue dashed curve) across the five non-SLy4 low-density branches as functions of $f=M/M_{\max}$; the logarithmic vertical scale emphasizes the monotonic convergence of $R_{\rm core}$. The curves cross near $f=0.75$, identifying the point beyond which the core becomes less sensitive than the total radius to the low-density branch and therefore supplies the more stable radial scale.}
    \label{fig:core_vs_fmax}
\end{figure*}

This behavior is governed by the proximity to the maximum-mass configuration rather than by the absolute mass. Because each low-density branch yields a slightly different maximum mass, between $2.301$ and $2.357\,M_\odot$ for the set considered here, we also compare stars at fixed $f\equiv M/M_{\max}$. Figure~\ref{fig:core_vs_fmax} shows that the core-radius spread falls monotonically from $490$~m at $f=0.60$ to $78$~m at $f=0.96$, a factor of $6.3$, while the total-radius spread decreases only from $377$ to $289$~m. The two curves cross near $f\simeq0.75$, which marks the transition between the two regimes identified above. The convergence of $R_{\rm core}$ is therefore a systematic property of the approach to the turning point and not a consequence of the particular fiducial calibration adopted here. We restrict the comparison to $f\le0.96$ because $dM/d\epsilon_c\to0$ near the maximum makes the determination of the central density ill-conditioned above that value.

Two caveats accompany this conclusion. First, the fixed-$f$ comparison is a diagnostic of the mechanism rather than an observational statement: it compares stars of different masses, and $M_{\max}$ is not measurable for an individual object. The fixed-mass comparison of Fig.~\ref{fig:core_universality} is the one that connects directly to PSR~J0030+0451 and PSR~J0740+6620. Second, although the core radius is the more stable of the two quantities at high mass, the observable total radius still spreads by $\sim0.3$~km, because $\Delta R_{\rm outer}$ re-injects the low-density dependence. A single measured radius therefore cannot isolate the core-dominated scale; resolving it requires sub-kilometre, roughly $\pm500$~m, radius precision. Finally, the SLy4 baseline used for our fiducial sequence is the stiffest branch considered and has the largest $\Delta R_{\rm outer}$, which is why its total radii sit above those of the BSk and Skyrme group; a softer unified low-density branch would lower them by up to $\sim1$~km without altering the core-dominated behavior.

\subsection{Tidal Deformability and Multimessenger Consistency}
\label{sec:tidal}

\begin{figure*}[h]
    \centering
    \includegraphics[width=\textwidth]{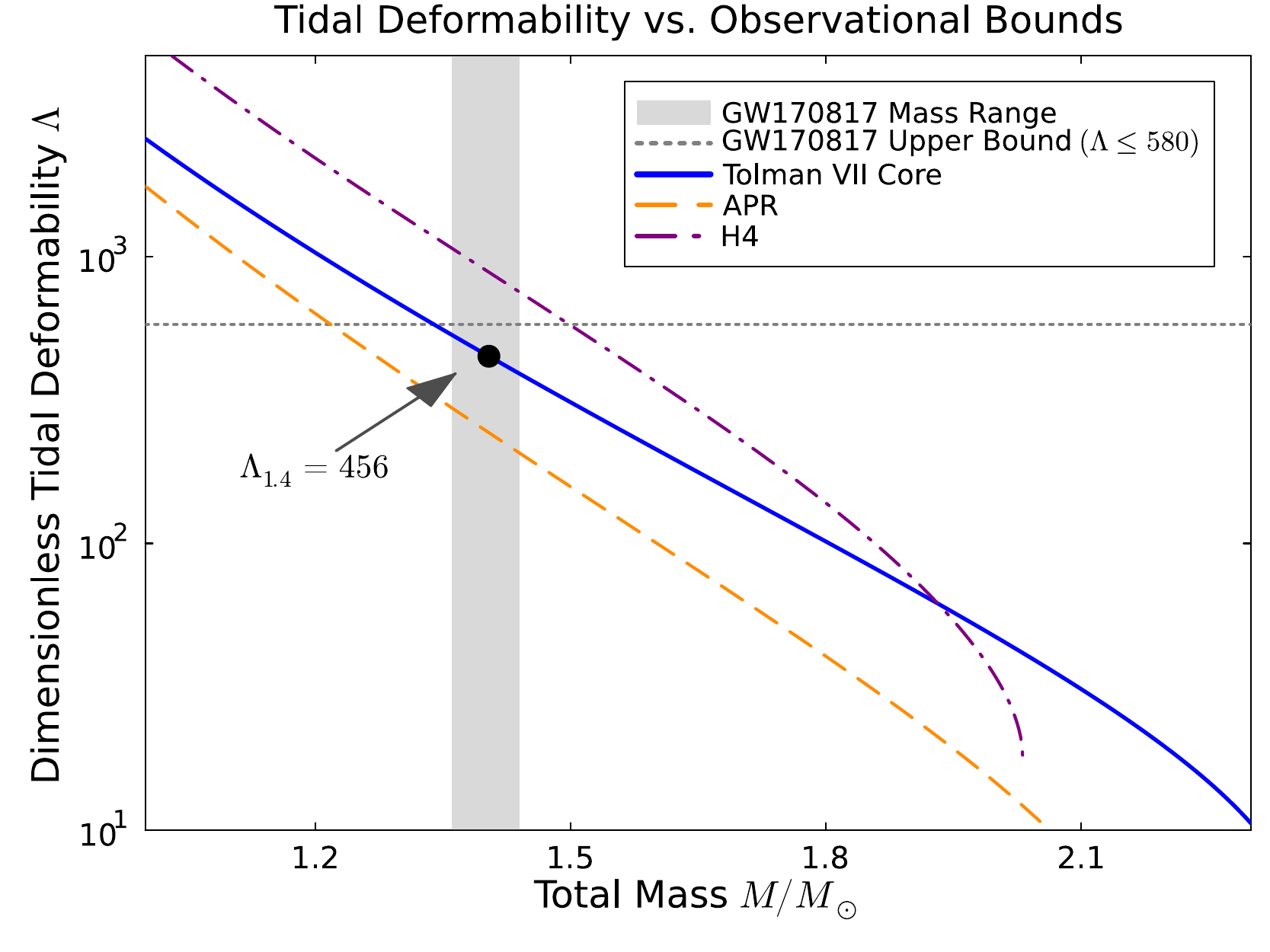}
    \caption{Dimensionless tidal deformability $\Lambda$ as a function of total stellar mass for APR, H4, and the present Tolman-VII-inspired stiff-core EOS. APR gives smaller radii and lower tidal deformabilities, H4 gives larger radii and larger tidal responses, and the present model lies between them at $1.4\,M_\odot$ while developing a distinct stiff-core high-mass behavior. The shaded band marks the GW170817-inspired $90\%$ confidence upper bound, $\Lambda_{1.4} \lesssim 580$; the fiducial stiff-core value, $\Lambda_{1.4} \approx 456$, lies below this bound and therefore satisfies the multimessenger constraint.}
    \label{fig:lambda}
\end{figure*}

To assess multimessenger consistency, we connect the proposed stiff-core EOS with the gravitational-wave constraints derived from binary neutron star mergers, most notably GW170817 \cite{abbott/2018}. The comparison in Fig.~\ref{fig:lambda} is motivated by two related questions: whether the near-vertical high-mass mass-radius behavior found here remains compatible with gravitational-wave tidal constraints, and where the Tolman-VII-inspired stiff-core model lies within the broader landscape of realistic neutron-star EOS families. During the late stages of a binary inspiral, the tidal fields of the companion stars induce quadrupole deformations. The degree of this deformation is quantified by the dimensionless tidal deformability parameter,
\begin{equation}
\Lambda \equiv \frac{2}{3}k_2 \left(\frac{R_\ast c^2}{GM}\right)^5,
\end{equation}
where $k_2$ is the second Love number. Because $\Lambda$ scales with the fifth power of the total stellar radius ($R_\ast^5$), it provides a sensitive, independent probe of the neutron-star radius and compactness. It therefore supplies a stringent consistency test for any EOS capable of producing massive pulsars with relatively large radii.

Figure~\ref{fig:lambda} compares the tidal deformability sequences of the present model with the APR and H4 EOS families, using the same LALSuite-based comparison data employed in the mass-radius analysis \cite{lalsuite,akmal/1998,lackey/2006,read/2009}. APR represents a comparatively softer nucleonic EOS and therefore produces smaller radii and lower tidal deformabilities. H4 represents a substantially stiffer conventional EOS, producing systematically larger radii and correspondingly larger tidal responses. The Tolman-VII-inspired stiff-core EOS occupies an intermediate but structurally distinct regime: at $1.4\,M_\odot$ it yields $\Lambda_{1.4}\approx456$, against $248$ for APR and $898$ for H4, so it is not as globally deformable as H4, yet it differs from APR by developing rapid high-density stiffening that reduces radial contraction in massive stars.

For our standard configuration ($s_{\rm tr}=0.40$ and $b=3.0\times10^{-4}\,(\mathrm{MeV\,fm^{-3}})^{-1}$), the integration yields $\Lambda_{1.4} \approx 456$, where $\Lambda_{1.4}$ denotes $\Lambda$ for a $1.4\,M_\odot$ star. This value remains compatible with the GW170817-inspired upper bound ($\Lambda_{1.4} \lesssim 580$ at $90\%$ confidence) despite the fact that the same EOS generates a near-vertical high-mass mass-radius branch. For the massive $2.08\,M_\odot$ configuration, the rapid stiffening keeps the star sufficiently compact to yield a low predicted deformability of $\Lambda_{2.08} \approx 34$. This provides a consistency requirement for any future high-mass binary neutron star merger observed by next-generation gravitational-wave detectors. This is an important result because strongly stiff EOS are often expected to overpredict tidal deformabilities by producing excessively large radii at $1.4\,M_\odot$.

The present model avoids this problem because the rapid relativistic stiffening develops primarily above approximately $2\epsilon_0$. Stars with $1.4\,M_\odot$ do not yet probe the deepest stiff-core regime, so their global deformability remains anchored by the lower- and intermediate-density sectors of the EOS. The same EOS can therefore remain moderately deformable at low mass while substantially reducing radial contraction at high mass, once the central energy density enters the rapidly stiffening relativistic core regime.

The comparison also clarifies the structural distinction among the EOS families. APR exhibits stronger compactification and a faster decrease in tidal deformability with increasing mass. H4 remains globally stiffer and more deformable across the sequence. The Tolman-VII-inspired model achieves a different balance by combining a moderate $1.4\,M_\odot$ deformability with a rapidly stiffening relativistic core at higher densities. Thus the near-vertical branch is not simply equivalent to a uniformly stiff EOS such as H4. Its defining feature is the onset of sufficiently rapid high-density stiffening above the transition density, which allows the high-density core to dominate the massive-star radius size without excessively enlarging $1.4\,M_\odot$ stars.

The precise APR and H4 tidal-deformability values should likewise be interpreted as representative benchmarks. Minor implementation choices can shift the numbers slightly, but the qualitative structural distinctions remain robust: APR is comparatively compact and weakly deformable, H4 is globally stiff and strongly deformable, and the present Tolman-VII-inspired EOS combines a GW170817-compatible $1.4\,M_\odot$ deformability with suppressed high-mass radial contraction.

Tidal deformability therefore provides a crucial complementary observable. While NICER constrains the mass-radius structure, gravitational waves independently constrain the global deformability and overall compactness. The combined NICER and GW170817 consistency of the Tolman-VII-inspired EOS strengthens the interpretation that massive neutron stars may already be probing a regime of early relativistic stiffening in dense matter above several times nuclear saturation density.

\subsection{Parameter-Space Structure and Robustness of the Stiff-Core Regime}
\label{sec:sensitivity}

\begin{figure*}[h]
    \centering
    \includegraphics[width=\textwidth]{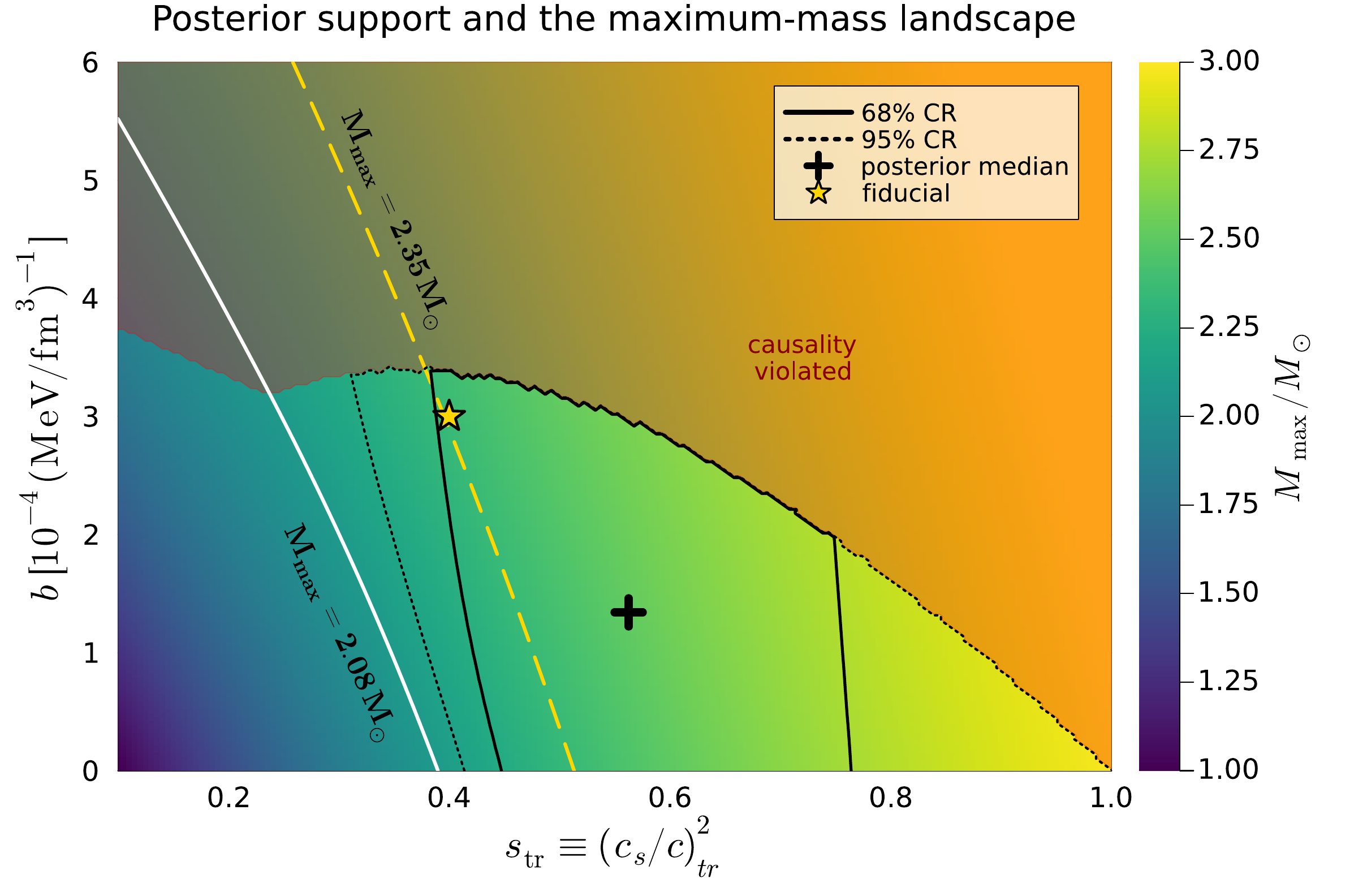}
    \caption{Posterior support and maximum-mass landscape for the piecewise SLy4-quadratic EOS. Background colors encode the maximum mass $M_{\max}$, while the solid white and dashed yellow curves mark $M_{\max}=2.08\,M_\odot$ and $2.35\,M_\odot$, respectively. The orange region above the red boundary violates causality ($s_{\max}>1$). Solid and dotted black contours enclose the $68\%$ and $95\%$ Bayesian credible regions obtained from the combined NICER and GW170817 inference; they are terminated at the causal boundary, against which they are cut rather than closed by the data. The black cross marks the posterior median and the gold star the fiducial calibration used for the representative stellar sequence. Note that the $b$ coordinate of the cross is not an observational determination: the marginal posterior for $b$ is set by causality, and its $95\%$ upper limit differs by about $3\%$ from that of the admissible prior alone.}
    \label{fig:posterior_mmax}
\end{figure*}

The Bayesian analysis described in Sec.~\ref{sec:bayesian_method} determines
whether the near-vertical mass-radius behavior found above is demanded by
current data or merely produced by a finely tuned EOS calibration. Its
posterior support is summarized together with the maximum-mass landscape in
Fig.~\ref{fig:posterior_mmax}.

The quadrature yields the log evidence $\ln\mathcal{Z}_{\rm quad}=-2.6754$
and the marginalized constraints
\begin{equation}
\begin{aligned}
s_{\rm tr} &= 0.563^{+0.196}_{-0.151}, \\
b_{\rm med} &= 1.341\times10^{-4}\ {\rm MeV}^{-1}{\rm fm}^{3}, \\
b &< 3.017\times10^{-4}\ {\rm MeV}^{-1}{\rm fm}^{3}\quad (95\%).
\end{aligned}
\end{equation}
These are posterior constraints, not manually selected benchmark
parameters, and the two parameters are constrained in qualitatively
different ways. For $s_{\rm tr}$ the data are informative but only modestly
so: the $68\%$ interval narrows from a width of $0.432$ under the admissible
prior alone to $0.347$ in the posterior, and the median shifts from $0.487$
to $0.563$. For $b$ they are not: the $95\%$ upper limit moves only from
$3.117\times10^{-4}$ under that prior to $3.017\times10^{-4}$ in the posterior,
so the marginal for $b$ is fixed by causality rather than measured by the
observations. The total information gain of the posterior relative to the
prior is $0.233$ nats. This asymmetry is the
statistical counterpart of the central result of this work: $b$ acts only
on the deep core, whereas $R_{1.34}$ and $\Lambda_{1.4}$ are set by the
region below it, so no combination of the present constraints can resolve
the quadratic stiffening.

The fiducial calibration used for the representative stellar sequence,
\begin{equation}
    (s_{\rm tr},b)=\left(0.40,\;3.0\times10^{-4}\,\mathrm{MeV}^{-1}\mathrm{fm}^3\right),
\end{equation}
lies inside the admissible posterior support. In $s_{\rm tr}$ it sits just
below the $68\%$ interval, at $0.40$ against a posterior median of $0.563$ and
a lower bound of $0.412$, so the representative sequence is calibrated
conservatively with respect to the inference. In $b$ it sits at
$3.0\times10^{-4}\,\mathrm{MeV}^{-1}\mathrm{fm}^{3}$, just below the $95\%$
upper limit of $3.017\times10^{-4}$ and therefore essentially on the causal
boundary, which is what allows the sequence to reach the highest observed
pulsar masses.

The combined representation in Fig.~\ref{fig:posterior_mmax} also exposes
the macroscopic mechanics of the parameter plane. A softer transition
interface requires a larger quadratic stiffening coefficient to cross the
$2.08\,M_\odot$ and $2.35\,M_\odot$ thresholds, whereas a larger
$s_{\rm tr}$ reduces the additional stiffening required at higher density.
The credible regions are elongated along the direction of decreasing $b$ with
increasing $s_{\rm tr}$, with a posterior correlation coefficient of $-0.509$.
This is not an observational degeneracy: the same coefficient computed for the
admissible prior alone is also $-0.509$, so the shape is entirely set by the
geometry of the allowed region, in which large $b$ is permitted only at small
$s_{\rm tr}$. Of the $386$ sampled points on the $95\%$ contour, only $120$
belong to a genuine isodensity segment; the remaining points coincide with the
admissible-domain boundary. The corresponding count is $173$ of $309$ points
for the $68\%$ contour. The data therefore close the region only on its
low-$s_{\rm tr}$ side and do not alter the correlation at the reported
precision. The trade-off above remains a valid statement about the
maximum-mass contours, but the posterior does not probe it, because the
credible regions lie well above the $2.08\,M_\odot$ threshold where that
constraint is active. The fiducial calibration lies directly on the
$2.35\,M_\odot$ contour, illustrating why support for the most massive
observed pulsars occupies the stiff edge of the inferred parameter space.

The same map makes the causal restriction explicit. Increasing both
$s_{\rm tr}$ and $b$ eventually drives the central sound speed of the
maximum-mass configuration above the speed of light, producing the excluded
upper region. The likelihood vanishes beyond that boundary, so the credible
regions shown in Fig.~\ref{fig:posterior_mmax}
contain only causal, physically viable sequences; in the $b$ direction they
are terminated by it rather than closed by the data.

To quantify whether the additional quadratic degree of freedom is
required, we repeated the inference for the linear reference model of
Sec.~\ref{sec:linear_eos} ($b\equiv0$, single parameter $s$, the same
likelihood and a prior uniform over its own admissible region), obtaining
$\ln\mathcal{Z}_{\rm lin}=-2.5612$ against
$\ln\mathcal{Z}_{\rm quad}=-2.6754$. The resulting Bayes factor is
\begin{equation}
\ln\mathcal{B} = -0.1142,
\end{equation}
so the negative sign corresponds to a marginal preference for the linear
constant-sound-speed core. Its magnitude lies well within the inconclusive
range of the Jeffreys scale ($|\ln\mathcal{B}|<1$). Refining the
$(s_{\rm tr},b)$ grid from $60\times60$ to $180\times180$ changes
$\ln\mathcal{B}$ by less than $0.005$. Refining the central-density sweep from
$100$ to $200$ points leaves the result unchanged at the quoted precision,
whereas a coarser sweep of $50$ points is not sufficient and shifts
$\ln\mathcal{B}$ by approximately $0.07$.

This is an explicit manifestation of the structural degeneracies inherent to
current radius measurements: within the broad observational uncertainties,
the two models achieve comparable maximum likelihoods, so the additional
degree of freedom introduced by $b$ neither improves the fit appreciably nor
is strongly penalized. Current multimessenger data therefore allow, but do
not yet require, progressive stiffening above $2\epsilon_0$ within this
parametrization.

The choice of prior for $b$ is not a matter of convention here, and it is
worth making the reason explicit. Because the likelihood is compactly
supported in $b$ by causality alone, the evidence of the quadratic model
scales as the inverse of the prior width once that width exceeds the causal
boundary, so $\ln\mathcal{B}$ can be driven to any negative value simply by
enlarging the rectangle. With a rectangular prior
$b\sim\mathcal{U}(0,B)$, the reliable calculations give
$\ln\mathcal{B}=-1.142$ for $B=6\times10^{-4}$ and $-1.835$ for
$B=1.2\times10^{-3}\ \mathrm{MeV^{-1}fm^{3}}$, with corresponding quadratic
log evidences $-3.703$ and $-4.396$. A formal extrapolation to
$B=3\times10^{-4}$ gives $\ln\mathcal{B}=-0.449$, but this value is not used
because that box truncates genuine support: at low $s_{\rm tr}$ the causal
boundary extends to approximately $3.8\times10^{-4}$. This is the Lindley
effect acting on a parameter that the data do not constrain, not a preference
emerging from the observations. Restricting the prior to the admissible region
of each model fixes its volume by physics and removes the arbitrary dependence
on the enclosing width of $b$.

None of these conclusions depends on the detailed parametrization of the
likelihood. Repeating the calculation under four variants, replacing the
two-piece normal for the PSR J0740+6620 radius by a symmetric Gaussian of
the same mean width, adopting the Dittmann et al. \cite{dittmann/2024}
reduction in place of Salmi et al. \cite{salmi/2024}, replacing the
flat-topped GW170817 constraint by a Gaussian reproducing the same $90\%$
interval, and removing the tidal term altogether, gives
$-0.154\le\ln\mathcal{B}\le+0.072$, always within the inconclusive band. The
median of $s_{\rm tr}$ ranges from $0.512$ to $0.573$, a spread of $0.061$
compared with a typical $1\sigma$ uncertainty of approximately $0.175$. The
$95\%$ upper limit on $b$ is $3.051\times10^{-4}\,{\rm MeV}^{-1}{\rm fm}^{3}$
for all variants except the Dittmann et al.\ radius reduction, for which it is
$2.949\times10^{-4}\,{\rm MeV}^{-1}{\rm fm}^{3}$. The absolute evidences shift
by up to $6.6$ nats for the Gaussian tidal likelihood, but these normalization
offsets are common to both models and cancel in $\ln\mathcal{B}$. Comparing
the baseline result, $\ln\mathcal{B}=-0.109$, with the calculation without
GW170817, $-0.112$, shows that the present tidal term contributes negligibly
because the predicted $\Lambda_{1.4}$ lies inside its flat plateau.

The transition at $2\epsilon_0$ should be understood as a controlled modeling
choice rather than as a universal microscopic boundary. Conventional
nucleonic descriptions such as SLy4 become increasingly uncertain in the
few-times-saturation regime, where many-body repulsion, hyperonic channels,
hadron-quark crossover, or quarkyonic dynamics may alter the pressure response
\cite{lackey/2006, masuda/2013, kojo/2015, mclerran/2019}. The value
$\epsilon_{\rm tr}=2\epsilon_0$ is therefore a choice of analysis rather than a
continuous parameter of the inference, but it is not an arbitrary one. To test
the sensitivity to this choice, we treat the transition density as a discrete
model index and repeat the full grid inference at
$\epsilon_{\rm tr}/\epsilon_0=1.5$, $2.0$, $2.5$ and $3.0$. The corresponding
quadratic log evidences are $\ln\mathcal{Z}=-4.96$, $-2.68$, $-3.19$ and
$-3.97$. With equal prior weight, the evidence is maximized at $2\epsilon_0$,
disfavouring $1.5\epsilon_0$ by $2.28$ nats and
$3.0\epsilon_0$ by $1.29$ nats, while $2.5\epsilon_0$ remains indistinguishable at
$0.51$ nats. The early transition is disfavoured because it
drives $\Lambda_{1.4}$ to $608$, above the GW170817 credible interval, while the
late one lowers $R_{1.4}$ to $11.8$~km. Over the range not disfavoured by the
data, $2.0\le\epsilon_{\rm tr}/\epsilon_0\le3.0$, the median of $s_{\rm tr}$ moves from
$0.558$ to $0.695$, a systematic smaller than its own $68\%$ uncertainty, and
$\ln\mathcal{B}$ remains between $-0.11$ and $-0.49$, inconclusive throughout.
The decomposition of Sec.~\ref{sec:radius_decomp} is nonetheless defined by
this threshold, and the fractions quoted there should be read as diagnostics
of the radial partition rather than as observables; what does not depend on
the threshold is the mass scaling of the outer correction established in
Sec.~\ref{sec:analytical}.

Across the full transition scan, the posterior median of $s_{\rm tr}$ ranges
from $0.253$ at $1.5\epsilon_0$ to $0.695$ at $3\epsilon_0$. The earliest transition
therefore gives $s_{\rm tr}<1/3$. Statements that the transition sound speed is
favored above the conformal value apply to the adopted $2\epsilon_0$ model and to
later transitions, not universally across the scan; the $1.5\epsilon_0$ case is,
however, the one most strongly disfavored by the data.

We therefore conclude that NICER- and GW170817-compatible near-vertical mass-radius behavior appears naturally when the EOS develops sufficiently rapid relativistic stiffening above approximately $2\epsilon_0$. The increasing control exerted by the high-density core over the massive-star radius size is not a fine-tuned artifact of one particular parametrization; it is fully consistent with the causal Bayesian multimessenger inference within this class of models.

\subsection{Sound-Speed Bounds and Ultra-Massive Compact Objects}
\label{sec:sound_speed_bounds}

The GW190814 gravitational-wave event included a secondary compact object with a mass of $2.50$-$2.67\,M_{\odot}$ \cite{abbott/2020a}. Determining whether such an object is an ultra-massive neutron star or a low-mass black hole requires understanding how much pressure dense matter can generate at supranuclear density while remaining causal.

A useful diagnostic is the dimensionless local sound-speed squared, $s=(c_s/c)^2=dP/d\epsilon$. At asymptotically high densities, perturbative QCD predicts that cold quark matter approaches the conformal value $s\rightarrow 1/3$, corresponding to an ultrarelativistic weakly interacting plasma \cite{kurkela/2010, annala/2020}. This conformal benchmark is an asymptotic limit at baryon chemical potential $\mu_B\to\infty$, not a strict bound at neutron-star densities; intermediate-density matter can therefore exceed $1/3$ without violating causality, as discussed above in terms of the trace anomaly.

For our fiducial calibration, the trace anomaly along the core decreases from $\Delta\simeq0.28$ at the transition to $\Delta\approx-0.18$ at the central density of the maximum-mass configuration ($\epsilon_c\simeq7.94\,\epsilon_0$), crossing zero near $\epsilon\approx4.9\,\epsilon_0$, where $P/\epsilon$ exceeds the conformal value $1/3$ and reaches $0.51$ at the center. This places the model in the negative-trace-anomaly regime well before the central density is attained and violates the trace-anomaly positivity condition considered by Fujimoto \textit{et al.}~\cite{fujimoto/2022}; within the present phenomenological construction, however, the binding requirement is causality. Directly evaluating Eq.~\eqref{eq:quad_cs} for the fiducial configuration ($s_{\rm tr}=0.40$ and $b=3.0\times10^{-4}\,({\rm MeV\,fm}^{-3})^{-1}$), the maximum-mass star reaches $M_{\max}=2.357\,M_{\odot}$ at $\epsilon_c\simeq1190.8$~MeV~fm$^{-3}$ ($7.94\,\epsilon_0$), where $s_{\max}=0.935$. The sequence therefore remains causal, although the margin is narrow: for this fixed value $s_{\rm tr}=0.40$, supporting the mass of PSR J0952-0607 requires a sound speed close to the causal bound. Across the full parameter plane, a larger transition sound speed can provide the same maximum-mass support with a smaller quadratic contribution.

The same mechanism becomes more restrictive in the ultra-massive regime. The maximum-mass landscape in Fig.~\ref{fig:posterior_mmax} shows that extending the nonrotating sequence toward a $2.6\,M_{\odot}$ star requires the high-stiffness side of the $(s_{\rm tr},b)$ plane, but this support can be obtained through different combinations of the two parameters: a larger $s_{\rm tr}$ permits a smaller quadratic coefficient, whereas a lower $s_{\rm tr}$ requires stronger high-density stiffening and approaches the causal boundary more closely. Configurations adjacent to that boundary do drive the massive-star core toward $s(\epsilon)\simeq1$, but near-causal sound speeds are not required throughout the entire $2.6\,M_{\odot}$-capable region. Thus, if the GW190814 secondary is indeed a neutron star within this class of nonrotating causal EOS models, the EOS must become very stiff above $2\epsilon_0$, although it need not become nearly causal immediately at the transition. If dense matter cannot provide such strong pressure support at intermediate densities, stable neutron-star configurations in the $2.5$-$2.7\,M_{\odot}$ range become increasingly difficult to realize.

This inference should be understood as model dependent, not as a definitive identification of the GW190814 secondary. Alternative possibilities remain open, including rapid uniform rotation, differential rotation in transient configurations, anisotropic stresses, hybrid-star configurations, strong phase transitions, exotic strongly interacting matter, or the low-mass black-hole interpretation. The purpose of the present discussion is not to exclude these scenarios, but to show that, within the causal relativistic stiff-core framework explored here, reaching the mass-gap regime naturally pushes the EOS toward extreme stiffness.

The GW190814 secondary is not an isolated case. A growing set of compact-object candidates with masses between the most massive neutron stars and the least massive black holes has been compiled by de S\'a \textit{et al.}~\cite{desa/2022}, and for many of these candidates the available data do not uniquely establish whether the object is a neutron star or a black hole. The question addressed here: how stiff dense matter must be to support such masses within a causal description, therefore extends beyond a single event and applies to any member of this population that is ultimately identified as a neutron star.

The same conclusion applies to maximum masses inferred statistically from the observed population. Values near $2.5$--$2.6\,M_\odot$~\cite{rocha/2023} lie beyond the reach of the present fiducial causal sequence, which terminates at $M_{\max}=2.357\,M_\odot\simeq2.4\,M_\odot$ before $s_{\max}$ exceeds unity.

\section{Discussion and Conclusions}
\label{sec:conclusions}

The interior composition of neutron stars and the behavior of matter beyond nuclear saturation density ($\rho_0$) remain central puzzles in multimessenger astrophysics. The conventional radius paradigm is most naturally understood as a statement about stars near $1.4\,M_\odot$, for which the radius is expected to correlate strongly with low-to-intermediate-density matter, including the crust and outer core. This work does not dispute that interpretation. Instead, we have argued that the same hierarchy can change for massive neutron stars, $M\gtrsim 2\,M_{\odot}$, whose central densities probe the regime in which relativistic high-density stiffening becomes structurally important.

Recent NICER measurements sharpen this distinction. Published analyses infer overlapping radius credible intervals for the lower-mass pulsar PSR J0030+0451 and the massive pulsar PSR J0740+6620. The near-equality of their radii, despite the large mass difference, suggests that radial compression is reduced as the star enters the high-density regime. In the interpretation developed here, this behavior arises when the EOS stiffens rapidly above approximately $2\epsilon_0$, causing the leading mass dependence of $R_\ast(M\gtrsim 2\,M_{\odot})$ to become increasingly controlled by the relativistic core, even though $R_\ast(M\sim 1.4\,M_{\odot})$ may still reflect intermediate-density physics.

The structural evidence comes from decomposing the total radius into its high-density-core and outer-layer contributions. At $2.08\,M_\odot$ the core radius, defined at the $\epsilon_{\rm tr}=2\epsilon_0$ interface, changes by only $\sim160$~m across the five non-SLy4 low-density branches spanning a factor of two in pressure at the transition, half the corresponding $326$~m variation of the total radius, whereas at $1.4\,M_\odot$ the core is the more branch-sensitive of the two and is partially compensated by the outer layer. The importance of the core for setting the radius is therefore increasingly large as the star grows more massive. Comparing stars at fixed fractions of their own maximum mass shows that this convergence is systematic: the core-radius spread falls monotonically from $490$~m at $M/M_{\max}=0.60$ to $78$~m at $0.96$, while the total-radius spread decreases more mildly from $377$ to $289$~m. The near-vertical branch and the core-dominated radius size are thus properties of the stiff relativistic core itself, not artifacts of the particular low-density model adopted. Consistently, the leading-order thin-crust relation used in the analytical argument reproduces the numerically integrated physical crust thickness to $\lesssim15\%$, improving toward the exact result precisely in the massive-star regime.

By anchoring a standard SLy4 low-density branch to an analytically motivated, parameterized Tolman VII core, we showed that rapid quadratic stiffening near $2\epsilon_0$ can account for the weak radius variation along the sequence. Our piecewise numerical integrations of the TOV equations reproduce a near-vertical mass-radius ($M$--$R_\ast$) trajectory compatible with current NICER constraints. The Bayesian inference, summarized together with the maximum-mass landscape in Fig.~\ref{fig:posterior_mmax}, gives $\ln\mathcal{Z}_{\rm quad}=-2.6754$ and marginalized constraints $s_{\rm tr}=0.563^{+0.196}_{-0.151}$ and $b<3.017\times10^{-4}\,{\rm MeV}^{-1}{\rm fm}^{3}$ at $95\%$ credibility, with all three admissibility filters imposed on every sequence. For the fiducial choice $s_{\rm tr}=0.40$, reaching the mass of PSR J0952-0607 requires $b$ close to the causal boundary and correspondingly a central sound speed close to the causal bound; within the broader family, a larger $s_{\rm tr}$ reduces the quadratic stiffening required. Thus the conclusion is not that the true crust should be neglected, but that for sufficiently massive neutron stars the stiff relativistic core can become the leading contributor to $R_\ast$ while the outer low-density layer supplies a smaller correction.

The comparison with standard APR- and H4-like EOS families further sharpens this interpretation. APR follows the expected compactifying behavior, and H4, despite its larger radii and comparatively stiff character, still bends leftward as the mass increases. The Tolman-VII-inspired sequence instead remains substantially more vertical in the massive-star regime. This demonstrates that the relevant structural ingredient is not stiffness in a generic sense, but the early and rapid growth of the high-density pressure response above approximately $2\epsilon_0$. In that regime the relativistic core behaves as an extended, comparatively rigid pressure-supporting region, producing weaker compactification than conventional EOS families even when those families are themselves relatively stiff.

The tidal-deformability comparison leads to the same conclusion from an independent multimessenger observable. APR is more compact and less deformable, H4 is globally stiffer and more deformable, while the present model remains intermediate at $1.4\,M_\odot$ with $\Lambda_{1.4}\approx456$, against $248$ for APR and $898$ for H4. Thus the stiff-core sequence is not simply an H4-like uniformly stiff EOS shifted into the NICER bands. Rather, it combines GW170817-compatible deformability at $1.4\,M_\odot$ with rapid high-density stiffening that becomes the leading structural influence only in the massive-star regime. This joint behavior allows the model to satisfy tidal constraints without losing the near-vertical high-mass mass-radius branch.

The physical implications of such early stiffening remain significant, but the null-hypothesis test is equally important. Within the adopted $\epsilon_{\rm tr}=2\epsilon_0$ model, the posterior median $s_{\rm tr}=0.563$ lies above the conformal value $1/3$, indicating a stiff transition into the high-density core. This statement is conditional on the transition choice: the disfavored $1.5\epsilon_0$ model instead gives $s_{\rm tr}=0.253<1/3$. The coefficient $b$, by contrast, is bounded by causality rather than measured, with its $95\%$ upper limit changing by only about $3\%$ from the admissible prior. The evidence comparison against a constant-sound-speed core gives $\ln\mathcal{Z}_{\rm lin}=-2.5612$ versus $\ln\mathcal{Z}_{\rm quad}=-2.6754$, and hence $\ln\mathcal{B}=-0.1142$. The negative sign represents a marginal preference for the simpler linear model, but the magnitude remains far inside the inconclusive range. Present radius error bars are therefore still wide enough that neither description can be selected. Within this likelihood, breaking the degeneracy requires reducing the radius uncertainty on the massive branch to roughly $\sim\pm500\,{\rm m}$, so that the residual separation between the linear and quadratic high-mass sequences is resolved by the data. EOS scenarios with strong softening from exotic phase transitions, such as hyperonization or weakly interacting quark matter, remain constrained because they typically reduce radii or maximum masses unless followed by additional stiffening. In this interpretation, matter in the deep core may already be in a strongly interacting, highly relativistic regime, but sharper radius and tidal measurements are required to establish this behavior statistically.

Future gravitational wave detections by the LIGO/Virgo/KAGRA network and continued X-ray profiling by NICER will further constrain the allowable parameter space for the high-density EOS, including the strong high-density stiffening needed if mass-gap candidates such as the GW190814 secondary are neutron stars. Within the assumptions explored here, the current data motivate a reevaluation of the traditional separation between ``radius physics'' and ``maximum-mass physics'': in the massive-star regime, both can become strongly coupled through the behavior of the high-density EOS.

Finally, as discussed in K\"opp \textit{et al.} \cite{kopp/2026}, modified theories of gravity or anisotropic pressures can also yield extended radii and distinct mass-radius correlations. Our results demonstrate, however, that near-vertical mass-radius sequences can emerge purely from causal, nonperturbative thermodynamic stiffening within standard General Relativity, providing a conservative microphysical baseline against which more exotic mechanisms must be compared.

\appendix

\section{Calculation of Tidal Deformability and the Second Love Number}
\label{app:tidal}

The dimensionless tidal deformability $\Lambda$ is a crucial macroscopic observable characterizing a neutron star's response to the tidal field of a companion in a binary system. It is related to the stellar compactness $C = GM/(R_\ast c^2)$ and the $l=2$ tidal Love number $k_2$ via
\begin{equation}
    \Lambda = \frac{2}{3} k_2 C^{-5}.
\end{equation}
To compute $k_2$, we evaluate the static, linear even-parity perturbations to the spherically symmetric stellar metric \cite{hinderer/2008, damour/2009}. Inside the star, the spatial radial part of the metric perturbation is characterized by a dimensionless function $y(r)$ (introduced by Hinderer \cite{hinderer/2008, hinderer/2010, postnikov/2010}), which satisfies the first-order non-linear Riccati differential equation:
\begin{equation}
    r \frac{dy}{dr} + y^2 + y e^{\lambda} \left[ 1 + 4\pi r^2 (P - \epsilon) \right] + r^2 Q = 0,
\end{equation}
where we have adopted geometric units ($G = c = 1$). The metric function $e^{\lambda}$ is given by $e^{\lambda} = \left(1 - 2m(r)/r\right)^{-1}$, and the source term $Q(r)$ is defined as:
\begin{equation}
    Q \equiv 4\pi e^{\lambda}
    \left(5\epsilon + 9P + (\epsilon + P)\frac{d\epsilon}{dP}\right)
    - \frac{6e^{\lambda}}{r^2} - \left(\frac{d\nu}{dr}\right)^2.
\end{equation}
with the radial derivative of the time-time metric component $\nu(r)$ given by the background TOV equations:
\begin{equation}
    \frac{d\nu}{dr} = 2 e^{\lambda} \left( \frac{m(r) + 4\pi r^3 P}{r^2} \right).
\end{equation}
In these units, the derivative $d\epsilon/dP$ is the inverse of the dimensionless sound-speed squared, $d\epsilon/dP=1/s=(c/c_s)^2$.

The integration of $y(r)$ proceeds outward from the stellar center ($r=0$) simultaneously with the TOV equations for mass $m(r)$ and pressure $P(r)$. The equation is regularized at the origin by imposing the central boundary condition:
\begin{equation}
    y(0) = 2.
\end{equation}

For models containing sharp phase transitions in which the energy density $\epsilon$ has a discontinuous jump $\Delta \epsilon$, the perturbation $y(r)$ must receive a corresponding discontinuity correction \cite{postnikov/2010}. In the present piecewise Tolman-VII-inspired framework, however, we explicitly enforce $C^0$ continuity in both $P(r)$ and $\epsilon(r)$ at the core-envelope matching interface ($\epsilon_{\rm tr}=2\epsilon_0$). Consequently, the function $y(r)$ remains strictly continuous throughout the stellar interior up to the surface $r=R_\ast$.

At the stellar surface ($P(R_\ast) = 0$), the interior perturbation must match smoothly to the exterior spacetime metric. Evaluating the integrated function at the surface yields $y_R \equiv y(R_\ast)$. The second Love number $k_2$ is then computed analytically from $y_R$ and the final compactness $C$ using the standard exterior matching formula \cite{hinderer/2008, postnikov/2010}:
\begin{equation}
\begin{aligned}
    k_2(C,y_R) ={}& \frac{8C^5}{5}(1-2C)^2
    \left[2-y_R+2C(y_R-1)\right] \\
    &\times \Big\{2C\left[6-3y_R+3C(5y_R-8)\right] \\
    &\quad +4C^3\left[13-11y_R+C(3y_R-2)
    +2C^2(1+y_R)\right] \\
    &\quad +3(1-2C)^2\left[2-y_R+2C(y_R-1)\right]
    \ln(1-2C)\Big\}^{-1}.
\end{aligned}
\end{equation}
This surface matching condition guarantees that the internal microphysics encoded dynamically within $y_R$ translates accurately to the macroscopic tidal deformability $\Lambda$ probed by gravitational-wave observatories.

\backmatter

\bmhead{Acknowledgments}
We acknowledge fruitful discussions and the exchange of foundational ideas with Manuel Malheiro, whose contributions helped initiate this project. This work is a tribute to his memory. RVL was supported by INCT-FNA (Instituto Nacional de Ci\^encia e Tecnologia, F\'{\i}sica Nuclear e Aplica\c c\~oes), research Project No.~464898/2014-5, and acknowledges support from CAPES/FAPERJ/CNPq.

\bibliography{references}

\end{document}